\documentclass[pre,tighten]{revtex4}

\usepackage{amssymb,amsmath}

\usepackage{epsfig}

\usepackage{graphicx}

\newcommand{\dd}{\text{d}}

\begin{document}
\title{Mass transport of an impurity in a strongly sheared granular gas}
\author{Vicente Garz\'{o}\footnote[1]{Electronic address: vicenteg@unex.es}}
\affiliation{Departamento de F\'{\i}sica, Universidad de
Extremadura, E-06071 Badajoz, Spain}
\begin{abstract}
Transport coefficients associated with the mass flux of an impurity immersed in a granular gas under simple
shear flow are determined from the inelastic Boltzmann equation. A normal solution is obtained via a
Chapman-Enskog-like expansion around a local shear flow distribution that retains all the hydrodynamic orders in
the shear rate. Due to the anisotropy induced by the shear flow, tensorial quantities are required to describe
the diffusion process instead of the conventional scalar coefficients. The mass flux is determined to first
order in the deviations of the hydrodynamic fields from their values in the reference state. The corresponding
transport coefficients are given in terms of the solutions of a set of coupled linear integral equations, which
are approximately solved by considering the leading terms in a Sonine polynomial expansion. The results show
that the deviation of these generalized coefficients from their elastic forms is in general quite important,
even for moderate dissipation.

\noindent{\it Keywords}: transport processes/heat transfer (theory), granular matter, transport properties
(theory)

\end{abstract}

\date{\today}
\maketitle

\section{Introduction\label{sec1}}

The study of transport properties of an impurity or intruder immersed in a granular gas described by the
inelastic Boltzmann equation is quite an interesting problem. In particular, when the gas is in homogeneous
cooling state (HCS), the mass flux ${\bf j}_0$ for the impurity can be obtained by solving the Boltzmann
equation by means of the Chapman-Enskog expansion \cite{CC70} around the local version of the HCS. In the first
order of the expansion, the mass flux ${\bf j}_0$ is linear in the gradients of mole fraction, pressure and
temperature and in the external applied force. The corresponding transport coefficients are the diffusion
coefficient $D$, the pressure diffusion coefficient $D_p$, the thermal diffusion coefficient $D_T$ and the
mobility $\chi$. As in the elastic case, these coefficients are the solutions of a set of coupled integral
equations \cite{GD02,GMD06} which can be approximately solved by considering the leading terms in a Sonine
polynomial expansion. Explicit expressions for these transport coefficients have been obtained \cite{GD02,GMD06}
in terms of the coefficients of restitution and the parameters of the mixture (masses and sizes). These
analytical results show in general a good agreement with those obtained from computer simulations
\cite{BRCG00,LBD02,MG03,GM04}, even for strong dissipation.

However, the above HCS is not accessible experimentally. In order to keep a granular system fluidized, an
external energy supply is required. In some experimental situations, the gas is driven into rapid flow by the
presence of a shear field. In this case, a steady state is reached when viscous heating is exactly balanced by
the energy dissipation in collisions. This unbounded shear flow problem is usually referred to as the simple or
uniform shear flow (USF) and its study has received a great deal of attention in the past years, especially in
the case of monocomponent systems \cite{C90,G03}. Nevertheless, much less is known on transport in sheared
granular mixtures \cite{CH02,AL02,MG02,L04}. In particular, the understanding of mass transport in granular
shear flows is of practical interest, since, for instance, powders must frequently be mixed together before any
sort of process can begin. Due to the complexity of the general problem (binary sheared granular mixture), to
gain some insight one can first consider the special case of impurity or tracer particles, namely, a binary
mixture where the mole fraction of one of the components is negligible. The tracer or impurity problem is more
amenable to analytical treatment since tracer particles are directly enslaved to the granular gas and there are
fewer parameters than in a general binary mixture problem. Even in this limit case, the analysis of mass
transport in a strongly shearing granular gas is an intricate problem basically due to the anisotropy induced in
the system by the shear flow. For this reason, tensorial quantities $\{D_{ij}, D_{p,ij}, D_{T,ij}, \chi_{ij}\}$
are required to describe the mass transport process instead of scalar coefficients $\{D, D_{p}, D_{T}, \chi\}$.
The aim of this paper is to get the above tensors in the framework of the inelastic Boltzmann equation.

Some previous attempts have been carried out earlier in the case of the diffusion tensor $D_{ij}$, especially in
the self-diffusion problem \cite{self}. However, in all these studies the diffusion was observed in only one
direction, usually the direction parallel to the velocity gradient. The full diffusion tensor has been obtained
in granular gases using kinetic theory \cite{G02}, measured in simulations of rapid granular shear \cite{C97},
and even in some experiments of dense, granular shear flows in a two-dimensional Couette geometry \cite{UB04}.
Substantial work on granular diffusivity has been carried out by Hsiau and co-workers \cite{HS99}, who have
measured self-diffusion coefficients in a variety of granular systems. All these results clearly show that
diffusion is anisotropic and is significantly affected by inelasticity.

At a kinetic level, in the tracer limit one can assume that the velocity distribution function $f({\bf r}, {\bf
v},t)$ of the granular gas (excess component) obeys a closed (nonlinear) Boltzmann equation, while the velocity
distribution function $f_0({\bf r}, {\bf v},t)$ of the impurity satisfies a (linear) Boltzmann-Lorentz equation.
Both kinetic equations have been solved in the (pure) USF problem to get the rheological properties of the
mixture in the steady state \cite{MG02,G02}. Let us assume now that the system (granular gas plus impurity) is
in a state that deviates from the USF by {\em small} spatial gradients. In addition, we also assume that the
impurity and gas particles are subjected to the action of external forces. Since the gas is slightly perturbed
from the USF, the Boltzmann equation can be solved by expanding in small gradients around the (local) shear flow
distribution instead of the (local) HCS. This Chapman-Enskog-like expansion has been very recently used
\cite{L06,G06} to determine the heat and momentum fluxes to first order in the deviations of the hydrodynamic
field gradients from their values in the reference shear flow state. Once the state of the excess component is
well characterized, the goal here is to determine the mass transport of impurity by solving the
Boltzmann-Lorentz equation by means of a similar perturbation scheme, namely, a gradient expansion around the
corresponding shear flow distribution $f_0^{(0)}$ which applies for arbitrary values of the shear rate. In the
first order of the expansion, the tensors $D_{ij}$, $D_{p,ij}$, $D_{T,ij}$, and $\chi_{ij}$ are identified from
the mass flux ${\bf j}_0$. As was already pointed out in Refs.\ \cite{L06,G06}, an important point is that, for
general small deviations from the shear flow state, the zeroth-order distribution $f_0^{(0)}$ is not a
stationary distribution since the collisional cooling cannot be compensated {\em locally} for viscous heating.
This fact gives rise to new conceptual and practical difficulties not present in the results based on the
conventional Chapman-Enskog expansion. Due to these difficulties, the results will be restricted here to
particular perturbations for which steady-state conditions apply and so the (reduced) shear rate (which is the
relevant nonequilibrium parameter of the problem) is coupled to dissipation.

The plan of the paper is as follows. In Sec.\ \ref{sec2} we describe the problem we are interested in. The USF
state is analyzed and the nonzero elements of the pressure tensor for the gas and the impurity are obtained by
means of Grad's method \cite{MG02,G02,SGD04}. Section \ref{sec3} deals with the perturbation scheme used to
solve the Boltzmann-Lorentz equation for the impurity to first order in the spatial gradients. The results show
that the generalized transport coefficients $D_{ij}$, $D_{p,ij}$, $D_{T,ij}$, and $\chi_{ij}$ are the solutions
of a set of coupled linear integral equations. A Sonine polynomial approximation taking the shear flow
distribution $f_0^{(0)}$ as the weight function is applied in Sec.\ \ref{sec4} to solve these integral equations
and get explicit expressions for these transport coefficients. The details of the calculations are displayed
along several Appendices. The dependence of some of these transport coefficients on dissipation is illustrated
in the three-dimensional case, showing again that the influence of inelasticity on mass transport is quite
significant. Finally, the paper is closed in Sec.\ \ref{sec5} with some concluding remarks.

\section{Description of the problem\label{sec2}}

As already said, the aim of this work is to analyze mass transport of a dilute granular mixture subjected to
uniform shear flow (USF) in the tracer limit, i.e., when the mole fraction of one of the species is negligible.
In this case, the state of the granular gas (excess component) is not affected by the presence of the tracer
particles. At a kinetic theory level, this implies that the velocity distribution function of the excess
component obeys a closed nonlinear Boltzmann equation. Moreover, the mole fraction of the tracer particles is so
small that their mutual interactions can be neglected as compared with their interactions with the particles of
the excess component. As a consequence, the velocity distribution function of the tracer particles satisfies a
linear Boltzmann-Lorentz equation. This is formally equivalent to study an impurity or intruder in a dilute
granular gas, and this will be the terminology used here. Let us start by offering a short review on some basic
aspects of the inelastic Boltzmann equation and its solution in the USF state.

We consider a granular gas composed by smooth inelastic disks ($d=2$) or spheres ($d=3$) of mass $m$ and
diameter $\sigma$. The inelasticity of collisions among all pairs is accounted for by a {\em constant}
coefficient of restitution $\alpha$ ($0\leq \alpha \leq 1$) that only affects the translational degrees of
freedom of grains. We also assume that the particles feel the presence of an external conservative force ${\bf
F}$ (such as a gravity field). At low density,  the time evolution of the one-particle velocity distribution
function of the gas $f({\bf r}, {\bf v},t)$ is given by the inelastic Boltzmann equation \cite{GS95,BDS97}
\begin{equation}
\label{2.1} \left(\frac{\partial}{\partial t}+{\bf v}\cdot \nabla+\frac{{\bf F}}{m}\cdot
\frac{\partial}{\partial {\bf v}} \right)f({\bf r}, {\bf v},t)=J[{\bf v}|f(t),f(t)],
\end{equation}
where the Boltzmann collision operator $J[{\bf v}|f,f]$ is
\begin{eqnarray}
\label{2.2} J\left[{\bf v}_{1}|f,f\right]&=&\sigma^{d-1} \int \dd{\bf v}_{2} \int \dd\widehat{\boldsymbol
{\sigma }}\,\Theta (\widehat{{\boldsymbol {\sigma }}}
\cdot {\bf g})(\widehat{\boldsymbol {\sigma}}\cdot {\bf g})\nonumber\\
& & \times \left[ \alpha ^{-2} f({\bf r},{\bf v}_1')f({\bf r},{\bf v}_2',t) -f({\bf r},{\bf v}_1,t)f({\bf
r},{\bf v}_2,t)\right].
\end{eqnarray}
Here, $\widehat{\boldsymbol {\sigma}}$ is a unit vector along their line of centers, $\Theta $ is the Heaviside
step function, and ${\bf g}={\bf v}_{1}-{\bf v}_{2}$ is the relative velocity. The primes on the velocities
denote the initial values $\{{\bf v}_{1}^{\prime }, {\bf v}_{2}^{\prime }\}$ that lead to $\{{\bf v}_{1},{\bf
v}_{2}\}$ following a binary collision:
\begin{equation}
\label{2.3} {\bf v}_{1}^{\prime}={\bf v}_{1}-\frac{1}{2}\left(1+\alpha^{-1}\right) (\widehat{\boldsymbol
{\sigma}}\cdot {\bf g})\widehat{\boldsymbol {\sigma}}, \quad {\bf v}_{2}^{\prime }={\bf v}_{2}+\frac{1}{2}\left(
1+\alpha^{-1}\right) (\widehat{\boldsymbol {\sigma}}\cdot {\bf g})\widehat{\boldsymbol {\sigma}}
\end{equation}
The first five velocity moments of $f$ define the number density
\begin{equation}
\label{2.4} n({\bf r}, t)=\int \; \dd{\bf v}f({\bf r}, {\bf v},t),
\end{equation}
the flow velocity
\begin{equation}
\label{2.5} {\bf u}({\bf r}, t)=\frac{1}{n({\bf r}, t)}\int \; \dd{\bf v} {\bf v} f({\bf r},{\bf v},t),
\end{equation}
and the {\em granular} temperature
\begin{equation}
\label{2.6} T({\bf r}, t)=\frac{m}{d n({\bf r}, t)}\int \; \dd{\bf v}\; V^2 f({\bf r},{\bf v},t),
\end{equation}
where ${\bf V}({\bf r},t)\equiv {\bf v}-{\bf u}({\bf r}, t)$ is the peculiar velocity. The macroscopic balance
equations for density $n$, momentum $m{\bf u}$, and energy $\frac{d}{2}nT$ follow directly from Eq.\
({\ref{2.1}) by multiplying with $1$, $m{\bf v}$, and $\frac{1}{2}mv^2$ and integrating over ${\bf v}$:
\begin{equation}
\label{2.7} D_{t}n+n\nabla \cdot {\bf u}=0\;,
\end{equation}
\begin{equation}
\label{2.8} D_{t}{\bf u}+(mn)^{-1}\left(\nabla \cdot {\sf P}-n{\bf F}\right)=0\;,
\end{equation}
\begin{equation}
\label{2.9} D_{t}T+\frac{2}{dn}\left(\nabla \cdot {\bf q}+P_{ij}\nabla_j u_i\right) =-\zeta T\;,
\end{equation}
where $D_{t}=\partial _{t}+{\bf u}\cdot \nabla$ is the material time derivative. The microscopic expressions for
the pressure tensor ${\sf P}$, the heat flux ${\bf q}$, and the cooling rate $\zeta$ are given, respectively, by
\begin{equation}
{\sf P}({\bf r}, t)=\int \dd{\bf v}\,m{\bf V}{\bf V}\,f({\bf r},{\bf v},t),
 \label{2.10}
\end{equation}
\begin{equation}
{\bf q}({\bf r}, t)=\int \dd{\bf v}\,\frac{1}{2}m V^{2}{\bf V}\, f({\bf r},{\bf v},t), \label{2.11}
\end{equation}
\begin{equation}
\label{2.12} \zeta({\bf r}, t)=-\frac{1}{dn({\bf r},t)T({\bf r}, t)}\int\, \dd{\bf v} mV^2J[{\bf r},{\bf
v}|f(t)].
\end{equation}

Let us suppose now that an impurity or intruder of mass $m_0$ and diameter $\sigma_0$ is added to the gas. As
said before, the presence of the intruder does not perturb the state of the gas, so that its velocity
distribution function is still determined by the Boltzmann equation (\ref{2.1}). In addition, the macroscopic
flow velocity and temperature for the mixture composed by the gas plus the impurity are the same as those for
the gas, namely they are given by Eqs.\ (\ref{2.5}) and (\ref{2.6}), respectively. The velocity distribution
function $f_0({\bf r}, {\bf v},t)$ of the impurity satisfies the linear Boltzmann-Lorentz equation
\begin{equation}
\label{2.13} \left(\frac{\partial}{\partial t}+{\bf v}\cdot \nabla+\frac{{\bf F}_0}{m_0}\cdot
\frac{\partial}{\partial {\bf v}} \right)f_0({\bf r}, {\bf v},t)=J_0[{\bf v}|f_0(t),f(t)],
\end{equation}
where the collision operator $J_0[{\bf v}|f_0,f]$ is now
\begin{eqnarray}
\label{2.14} J_0\left[{\bf v}_{1}|f_0,f\right]&=&\overline{\sigma}^{d-1} \int \dd{\bf v}_{2} \int
\dd\widehat{\boldsymbol {\sigma}}\,\Theta (\widehat{{\boldsymbol {\sigma}}}
\cdot {\bf g})(\widehat{\boldsymbol {\sigma}}\cdot {\bf g})\nonumber\\
& & \times \left[ \alpha_0 ^{-2} f_0({\bf r},{\bf v}_1')f({\bf r},{\bf v}_2',t) -f_0({\bf r},{\bf v}_1,t)f({\bf
r},{\bf v}_2,t)\right].
\end{eqnarray}
Here, ${\bf F}_0$ denotes an external force acting on impurity, $\overline{\sigma}=(\sigma+\sigma_0)/2$ and
$\alpha_0$ ($0\leq \alpha_0 \leq 1$) is the coefficient of restitution for impurity-gas collisions. The
precollisional velocities are given by
\begin{equation}
{\bf v}_{1}^{\prime }={\bf v}_{1}-\frac{m}{m+m_0}\left( 1+\alpha _{0}^{-1}\right) (\widehat{{\boldsymbol {\sigma
}}}\cdot {\bf
g}_{12})\widehat{{\boldsymbol {\sigma }}} ,\nonumber\\
\end{equation}
\begin{equation}
 {\bf v}_{2}^{\prime }={\bf v}_{2}+\frac{m_0}{m+m_0}\left( 1+\alpha
_{0}^{-1}\right) (\widehat{{\boldsymbol {\sigma }}}\cdot {\bf g}_{12})\widehat{ \boldsymbol {\sigma}}.
\label{2.3bis}
\end{equation}
As shown in Ref.\ \cite{SD06}, the operator $J_0\left[{\bf v}|f_0,f\right]$ is the same as that of an {\em
elastic} impurity ($\alpha_0=1$) with an effective mass $m_0^*=m_0+(m_0+m)(1-\alpha_0)/(1+\alpha_0)$.

The number density for the impurity is
\begin{equation}
\label{2.15} n_0({\bf r},t)=\int\; \dd{\bf v}f_0({\bf r},{\bf v},t).
\end{equation}
The impurity may freely exchange momentum and energy with the particles of the gas and, therefore, these are not
invariants of the collision operator $J_0[{\bf v}|f_0,f]$. Only the number density $n_0$ is conserved, whose
continuity equation is directly obtained from Eq.\ (\ref{2.13})
\begin{equation}
\label{2.16} D_{t}n_0+n_0\nabla \cdot {\bf u}+\frac{\nabla \cdot {\bf j}_0}{m_0}=0\;,
\end{equation}
where ${\bf j}_0$ is the mass flux for the impurity, relative to the local flow ${\bf u}$,
\begin{equation}
{\bf j}_{0}=m_{0}\int \dd{\bf v}\,{\bf V}\,f_0({\bf r},{\bf v},t). \label{2.17}
\end{equation}
At a kinetic level, an interesting quantity is the local temperature of the impurity defined as
\begin{equation}
\label{2.18} T_0({\bf r}, t)=\frac{m_0}{d n_0({\bf r}, t)}\int \; \dd{\bf v}\, V^2 f_0({\bf r},{\bf v},t).
\end{equation}
This quantity measures the mean kinetic energy of the impurity. As will be shown later, the global temperature
$T$ and the temperature of the impurity $T_0$ are in general different, so that the granular energy per particle
is not equally distributed between both components of the system \cite{CH02,MG02}.

Let us start by describing the state of the system (gas plus impurity) in the (pure) USF. This idealized
macroscopic state is characterized by the absence of external forces (${\bf F}={\bf F}_0={\bf 0}$), constant
densities $n$ and $n_0$, a uniform temperature, and a linear velocity profile ${\bf u}={\sf a}\cdot {\bf r}$,
where the elements of the tensor ${\sf a}$ are $a_{ij}=a\delta_{ix}\delta_{jy}$, $a$ being the constant shear
rate. This linear velocity profile assumes no boundary layer near the walls and is generated by the Lee-Edwards
boundary conditions \cite{LE72}, which are simply periodic boundary conditions in the local Lagrangian frame
moving with the flow velocity. For elastic gases, the temperature grows in time due to viscous heating and so a
steady state is not possible unless an external (artificial) force is introduced \cite{GS03}. However, for
inelastic gases, the temperature changes in time due to the competition between two (opposite) mechanisms: on
the one hand, viscous (shear) heating and, on the other hand, energy dissipation in collisions. A steady state
is achieved when both mechanisms cancel each other and the fluid autonomously seeks the temperature at which the
above balance occurs. Under these conditions, in the steady state the balance equation (\ref{2.9}) becomes
\begin{equation}
\label{2.19} aP_{xy}=-\frac{d}{2}\zeta p,
\end{equation}
where $p=nT$ is the hydrostatic pressure. The balance equation (\ref{2.19}) shows the intrinsic connection
between the shear field and dissipation in the system. As a consequence, the shear flow state associated with
(\ref{2.19}) is inherently beyond the scope of the Navier-Stokes or Newtonian hydrodynamic equations
\cite{SGD04} since the collisional cooling (which is fixed by the mechanical properties of the particles) sets
the strength of the velocity gradient in the steady state. Furthermore, note that for given values of the shear
rate $a$ and the coefficient of restitution $\alpha$, the relation (\ref{2.19}) gives the temperature $T$ in the
steady state as a {\em unique } function of the density $n$. In fact, the reduced shear rate $a^*=a/\nu(n,T)$ is
{\em only} a function of the coefficient of restitution $\alpha$ in the steady state. Here, $\nu(n,T)$ is a
characteristic collision frequency given by
\begin{equation}
\label{2.19.1} \nu=\frac{\pi^{(d-1)/2}}{\Gamma(d/2)}\frac{8}{d+2} n\sigma^{d-1}\sqrt{\frac{T}{m}}.
\end{equation}

The relevant transport properties of the system in the steady USF are related to the pressure tensor ${\sf P}$
since ${\bf q}={\bf j}_0={\bf 0}$. Furthermore, one can introduce the partial pressure tensor ${\sf P}_0$
defined as
\begin{equation}
\label{2.19.2} {\sf P}_0=\int \dd{\bf v}\,m_0{\bf V}{\bf V}\,f_0({\bf V}).
\end{equation}
Explicit expressions for the nonzero elements of ${\sf P}$ and ${\sf P}_0$ have been recently obtained from the
Boltzmann equation by means of Grad's method \cite{G02}. A brief summary of these results is given in Appendix
\ref{appA}. The accuracy of this Grad's solution has been confirmed by comparison with Monte Carlo simulations
of the Boltzmann equation \cite{MG02,BRM97,AS05}, even for strong dissipation.

\section{Chapman-Enskog-like expansion: Transport properties
around USF\label{sec3}}

The main goal of this paper is to determine the mass flux for the impurity in the presence of USF. In that case,
let us assume that the USF state is disturbed by small spatial perturbations. The response of the system to
those perturbations gives rise to contributions to the mass flux, which can be characterized by generalized
transport coefficients. This Section is devoted to the evaluation of these coefficients.

In order to analyze this problem we have to start from the Boltzmann--Lorentz equation (\ref{2.13}) with a
general time and space dependence. Let ${\bf u}_s={\sf a}\cdot {\bf r}$ be the flow velocity of the {\em
undisturbed} USF state. In the {\em disturbed} state, however the true velocity ${\bf u}$ is in general
different from ${\bf u}_s$ since ${\bf u}={\bf u}_s+\delta {\bf u}$, $\delta {\bf u}$ being a small perturbation
to ${\bf u}_s$. As a consequence, the true peculiar velocity is now  ${\bf c}\equiv {\bf v}-{\bf u}={\bf
V}-\delta{\bf u}$, where ${\bf V}={\bf v}-{\bf u}_s$. In the Lagrangian frame moving with ${\bf u}_s$, the
Boltzmann--Lorentz equation can be written as
\begin{equation}
\label{3.1} \frac{\partial}{\partial t}f_0-aV_y\frac{\partial}{\partial V_x}f_0+\left({\bf V}+{\bf u}_s\right)
\cdot \nabla f_0+\frac{{\bf F}_0}{m_0}\cdot \frac{\partial}{\partial {\bf V}} f_0=J_0[{\bf V}|f_0,f],
\end{equation}
where here the derivative $\nabla f_0$ is taken at constant ${\bf V}$. The macroscopic balance equations
associated with this disturbed USF state follows from the general equations (\ref{2.8}), (\ref{2.9}), and
(\ref{2.16}) when one takes into account that ${\bf u}={\bf u}_s+\delta {\bf u}$. The result is
\begin{equation}
\label{3.2}
\partial_tn_0+{\bf u}_s\cdot \nabla n_0=-\nabla \cdot (n_0\delta {\bf u})-
\frac{\nabla \cdot {\bf j}_0}{m_0},
\end{equation}
\begin{equation}
\label{3.3}
\partial_t\delta {\bf u}+{\sf a}\cdot \delta {\bf u}+({\bf u}_s+\delta {\bf u})\cdot \nabla \delta {\bf u}=-
(mn)^{-1}\left(\nabla \cdot {\sf P}-n{\bf F}\right),
\end{equation}
\begin{equation}
\label{3.3.1} \frac{d}{2}n\partial_tT+\frac{d}{2}n({\bf u}_s+\delta {\bf u})\cdot \nabla T+aP_{xy}+\nabla \cdot
{\bf q}+{\sf P}:\nabla \delta {\bf u}=-\frac{d}{2}p\zeta,
\end{equation}
where the pressure tensor ${\sf P}$, the heat flux ${\bf q}$, the cooling rate $\zeta$, and the mass flux ${\bf
j}_0$  are defined by Eqs.\ (\ref{2.10}), (\ref{2.11}), (\ref{2.12}), and (\ref{2.17}), respectively, with the
replacement ${\bf V}\rightarrow {\bf c}$.

We assume that the deviations from the USF state are small. This means that the spatial gradients of the
hydrodynamic fields $A({\bf r},t)$ are small. As noted in Ref.\ \cite{GD02}, there is some flexibility in the
case of granular mixtures in the representation of the heat and mass fluxes since they can be defined in a
variety of equivalent ways depending on the choice of hydrodynamic gradients used. In fact, some care is
required in comparing transport coefficients in different representations using different independent gradients
for the driving forces. Here, as in previous works \cite{GD02,GMD06}, I take the concentration of impurity
$x_0=n_0/n$, the pressure $p$, the temperature $T$, and the local flow velocity $\delta {\bf u}$ as hydrodynamic
fields, i.e.,
\begin{equation}
\label{3.4} A({\bf r},t)\equiv \{x_0({\bf r},t), p({\bf r}, t), T({\bf r}, t), \delta {\bf u}({\bf r},t)\}.
\end{equation}
Since the system is strongly sheared, a solution to the Boltzmann--Lorentz equation (\ref{3.1}) can be obtained
by means of a generalization of the conventional Chapman-Enskog method \cite{CC70} where the velocity
distribution function is expanded about a {\em local} shear flow reference state in terms of the small spatial
gradients of the hydrodynamic fields relative to those of USF. This is the main new ingredient of the expansion.
This type of Chapman-Enskog-like expansion has been considered in the case of elastic gases to get the set of
shear-rate dependent transport coefficients \cite{GS03,LD97} in a thermostatted shear flow problem and it has
also been recently considered \cite{L06,G06} for inelastic gases.

In the context of the Chapman--Enskog method \cite{CC70}, we look for a {\em normal} solution of the form
\begin{equation}
\label{3.5} f_0({\bf r}, {\bf V},t)\equiv f_0(A({\bf r}, t), {\bf V}).
\end{equation}
This special solution expresses the fact that the space dependence of the reference shear flow is completely
absorbed in the relative velocity ${\bf V}$ and all other space and time dependence occurs entirely through a
{\em functional} dependence on the fields $A({\bf r}, t)$. Moreover, in the presence of external forces it is
necessary to characterize the magnitude of these forces relative to the gradients as well. Here, it is assumed
that the magnitude of the external forces ${\bf F}$ and ${\bf F}_0$ is of first order in perturbation expansion.
The functional dependence (\ref{3.5}) can be made local by an expansion of the distribution function in powers
of the hydrodynamic gradients:
\begin{equation}
\label{3.7} f_0(A({\bf r}, t), {\bf V}) =f_0^{(0)}({\bf V})+ f_0^{(1)}({\bf V})+\cdots,
\end{equation}
where the reference zeroth-order distribution function corresponds to the USF distribution function but taking
into account the local dependence of the concentration, pressure and temperature and the change ${\bf
V}\rightarrow {\bf V}-\delta{\bf u}({\bf r}, t)={\bf c}$. The successive approximations $f^{(k)}$ are of order
$k$ in the strength of the external forces as well as in the gradients of $x_0$, $p$, $T$, and $\delta {\bf u}$
but retain all the orders in the shear rate $a$. Here, only the first order approximation will be considered.

When the expansion (\ref{3.7}) is substituted into the definitions (\ref{2.10}), (\ref{2.11}), (\ref{2.12}), and
(\ref{2.17}), one gets the corresponding expansions for the fluxes and the cooling rate:
\begin{subequations}
\begin{equation}
\label{3.8} {\sf P}={\sf P}^{(0)}+{\sf P}^{(1)}+\cdots, \quad {\bf q}={\bf q}^{(0)}+{\bf q}^{(1)}+\cdots,
\end{equation}
\begin{equation}
\zeta=\zeta^{(0)}+\zeta^{(1)}+\cdots, \quad {\bf j}_0={\bf j}_0^{(0)}+{\bf j}_0^{(1)}+\cdots.
\end{equation}
\end{subequations}
Finally, as in the usual Chapman-Enskog method, the time derivative is also expanded as
\begin{equation}
\label{3.9}
\partial_t=\partial_t^{(0)}+\partial_t^{(1)}+\partial_t^{(2)}+\cdots,
\end{equation}
where the action of each operator $\partial_t^{(k)}$ is obtained from the hydrodynamic equations
(\ref{3.2})--(\ref{3.4}). These results provide the basis for generating the Chapman-Enskog solution to the
inelastic Boltzmann--Lorentz equation (\ref{3.1}).

\subsection{Zeroth-order approximation}

Substituting the expansions (\ref{3.7})--(\ref{3.9}) into Eq.\ (\ref{3.1}), the kinetic equation for $f_0^{(0)}$
is given by
\begin{equation}
\label{3.10}
\partial_t^{(0)}f_0^{(0)}-aV_y\frac{\partial}{\partial V_x}f_0^{(0)}=J_0[{\bf
V}|f_0^{(0)},f^{(0}],
\end{equation}
where use has been made of the fact that ${\bf F}_0$ is of first order in gradients. To lowest order in the
expansion the conservation laws give
\begin{equation}
\label{3.10.1}
\partial_t^{(0)}x_0=0,\quad
T^{-1}\partial_t^{(0)}T=p^{-1}\partial_t^{(0)}p= -\frac{2}{dp}a P_{xy}^{(0)}-\zeta^{(0)},
\end{equation}
\begin{equation}
\label{3.11}
\partial_t^{(0)}\delta u_i+a_{ij} \delta u_j=0.
\end{equation}
As shown in Refs.\ \cite{L06,G06}, for given values of $a$ and $\alpha$, the steady state condition (\ref{2.19})
establishes a mapping between the pressure $p$ and temperature $T$ so that every pressure corresponds to one and
only one temperature. Since the pressure $p({\bf r}, t)$ and temperature $T({\bf r}, t)$ are specified
separately in the {\em local} USF state, the viscous heating term $a|P_{xy}^{(0)}|$ only partially compensates
for the collisional cooling and so, the pressure and temperature depend on time. This implies that the
zeroth-order distributions for the gas $f^{(0)}$ and the impurity $f_0^{(0)}$ depend both on time through their
dependence on the pressure and temperature. Consequently, in general the reduced shear rate $a^*=a/\nu(p,T)$
depends on space and time so that, $a^*$ and $\alpha$ must be considered as independent parameters for general
infinitesimal perturbations around the USF state. This fact gives rise to conceptual and practical difficulties
not present in the case of elastic collisions \cite{LD97}.

Since $f_0^{(0)}$ is a normal solution, the time derivative in Eq.\ (\ref{3.10}) can be represented more
usefully as
\begin{eqnarray}
\label{3.12}
\partial_t^{(0)}f_0^{(0)}&=&\frac{\partial f_0^{(0)}}{\partial
x_0}\partial_t^{(0)} x_0+\frac{\partial f_0^{(0)}}{\partial p}\partial_t^{(0)} p+\frac{\partial
f_0^{(0)}}{\partial T}\partial_t^{(0)} T+\frac{\partial f_0^{(0)}}{\partial \delta
u_i}\partial_t^{(0)} \delta u_i\nonumber\\
&=&-\left(\frac{2}{d p}a P_{xy}^{(0)}+\zeta^{(0)}\right)\left(p\frac{\partial}{\partial
p}+T\frac{\partial}{\partial T}\right)f_0^{(0)}-a_{ij}\delta u_j
\frac{\partial}{\partial \delta u_i}f_0^{(0)}\nonumber\\
&=&-\left(\frac{2}{d p}a P_{xy}^{(0)}+\zeta^{(0)}\right)\left(p\frac{\partial}{\partial
p}+T\frac{\partial}{\partial T}\right)f_0^{(0)}+a_{ij}\delta u_j \frac{\partial}{\partial c_i}f_0^{(0)},
\end{eqnarray}
where in the last step we have taken into account that $f_0^{(0)}$ depends on $\delta {\bf u}$ only through the
peculiar velocity ${\bf c}$. Substituting (\ref{3.12}) into (\ref{3.10}) yields the following kinetic equation
for $f_0^{(0)}$:
\begin{equation}
\label{3.13} -\left(\frac{2}{dp}a P_{xy}^{(0)}+\zeta^{(0)}\right)\left(p\frac{\partial}{\partial
p}+T\frac{\partial}{\partial T}\right)f_0^{(0)} -ac_y\frac{\partial}{\partial c_x}f_0^{(0)}=J_0[{\bf
V}|f_0^{(0)},f^{(0}].
\end{equation}
To solve Eq.\ (\ref{3.13}) one needs to know the dependence of the momentum flux $P_{xy}^{(0)}$ on the pressure
$p$ and temperature $T$. A detailed study of this problem has been carried out in Ref.\ \cite{SGD04}. The first
nontrivial velocity moment of the distribution $f_0^{(0)}$ corresponds to the partial pressure tensor ${\sf
P}_{0}^{(0)}$ defined as
\begin{equation}
\label{3.14} {\sf P}_{0}^{(0)}=m_0\;\int \dd{\bf c}\,{\bf c}{\bf c}\,f_0^{(0)}({\bf c}).
\end{equation}
The temperature of the impurity $T_0(t)$ can be determined from the trace of ${\sf P}_{0}^{(0)}$. From Eq.\
(\ref {3.13}), one gets
\begin{equation}
\label{3.15} -\left(\frac{2}{dp}a P_{xy}^{(0)}+\zeta^{(0)}\right)\left(p\frac{\partial}{\partial
p}+T\frac{\partial}{\partial T}\right) P_{0,ij}^{(0)}+ a_{i\ell}P_{0,j\ell}^{(0)}+a_{j\ell}P_{0,i\ell}^{(0)}=
B_{ij},
\end{equation}
where $a_{ij}=a\delta_{ix}\delta_{jy}$ and
\begin{equation}
\label{3.16} B_{ij}=m_0\;\int \dd{\bf c}\,c_ic_j\;J_0[f_0^{(0)},f^{(0)}].
\end{equation}
A good estimate of the collisional moment $B_{ij}$ can be made by considering Grads's approximation \cite{G02}.
In this case, $B_{ij}$ is given by Eq.\ (\ref{c5}). The steady state solution of Eq.\ (\ref{3.15}) is also
displayed in Appendix \ref{appA}. However, in general the equations (\ref{3.15}) must be solved numerically to
get the dependence of the zeroth-order pressure tensor $P_{0,ij}^{(0)}(p,T)$ on pressure and temperature. The
behavior of the pressure tensors $P_{ij}^{(0)}$ and $P_{0,ij}^{(0)}$ near the (steady) USF state is studied in
Appendix \ref{appC}. In what follows, $P_{ij}^{(0)}$ and $P_{0,ij}^{(0)}$ will be considered as known functions
of $p$ and $T$.

\subsection{First-order approximation}

The analysis to first order in the gradients is similar to the one carried out in Ref.\ \cite{G06} for the
one-component case. Some details of this derivation are given in Appendix \ref{appB}. The distribution function
$f_0^{(1)}$ is of the form
\begin{equation}
\label{3.17} f_0^{(1)}={\boldsymbol {\cal A}}_{0}\cdot \nabla x_0+ {\boldsymbol {\cal B}}_{0}\cdot \nabla
p+{\boldsymbol {\cal C}}_{0}\cdot \nabla T+{\sf {\cal D}}_{0}:\nabla \delta {\bf u}+{\boldsymbol {\cal
E}}_{0}\cdot {\boldsymbol {\cal F}},
\end{equation}
where
\begin{equation}
\label{3.18} {\boldsymbol {\cal F}}={\bf F}_0-\frac{m_0}{m}{\bf F}.
\end{equation}
The vectors $\{{\boldsymbol {\cal A}}_{0}, {\boldsymbol {\cal B}}_{0}, {\boldsymbol {\cal C}}_{0}, {\boldsymbol
{\cal E}}_{0}\}$, and the tensor ${\sf {\cal D}}_{0}$ are functions of the true peculiar velocity ${\bf c}$.
They are the solutions of the following linear integral equations:
\begin{equation}
\label{3.19} -\left(\frac{2}{dp}a
P_{xy}^{(0)}+\zeta^{(0)}\right)\left(p\partial_p+T\partial_T\right){\boldsymbol {\cal A}}_{0}- a
c_y\frac{\partial}{\partial c_x}{\boldsymbol {\cal A}}_{0}-J_0[{\boldsymbol {\cal A}}_{0},f^{(0)}]={\bf A}_{0},
\end{equation}
\begin{eqnarray}
\label{3.20}  & &-\left(\frac{2}{d p}a
P_{xy}^{(0)}+\zeta^{(0)}\right)\left(p\partial_p+T\partial_T\right){\boldsymbol {\cal B}}_{0}-
\left(\frac{2a}{d}\partial_pP_{xy}^{(0)}+2\zeta^{(0)}+ a c_y\frac{\partial}{\partial c_x}\right){\boldsymbol
{\cal B}}_{0}\nonumber\\
& & -J_0[{\boldsymbol {\cal B}}_{0},f^{(0)}]={\bf
B}_{0}-\left[\frac{2aT}{dp^2}(1-p\partial_p)P_{xy}^{(0)}-\frac{T\zeta^{(0)}}{p} \right]{\boldsymbol {\cal
C}}_{0}+J_0[f_0^{(0)},{\boldsymbol {\cal B}}],
\end{eqnarray}
\begin{eqnarray}
\label{3.21}  & &-\left(\frac{2}{d p}a P_{xy}^{(0)}+\zeta^{(0)}\right)\left(p\partial_p+T\partial_T\right)
{\boldsymbol {\cal C}}_{0} -\left[\frac{2a}{d p}\left(1+T\partial_T\right)P_{xy}^{(0)}+\frac{1}{2}\zeta^{(0)}
\right.\nonumber\\
& & \left. + a c_y\frac{\partial}{\partial c_x}\right]{\boldsymbol {\cal C}}_{0}-J_0[{\boldsymbol {\cal
C}}_{0},f^{(0)}]={\bf C}_{0}+\left(\frac{2a}{d}\partial_TP_{xy}^{(0)}-\frac{p\zeta^{(0)}}{2T}
\right){\boldsymbol {\cal B}}_{0}\nonumber\\
& & +J_0[f_0^{(0)},{\boldsymbol {\cal C}}],
\end{eqnarray}
\begin{eqnarray}
\label{3.22}  & &-\left(\frac{2}{d p}a P_{xy}^{(0)}+\zeta^{(0)}\right)\left(p\partial_p+T\partial_T\right){\cal
D}_{0,\ell j} - a c_y\frac{\partial}{\partial c_x}{\cal D}_{0,\ell
j}-a\delta_{\ell y}{\cal D}_{0,xj}\nonumber\\
& & -J_0[{\cal D}_{0,\ell j},f^{(0)}]=D_{0,\ell j} +\zeta_{u,\ell j}
\left(p\partial_p+T\partial_T\right)f^{(0)}+J_0[f_0^{(0)},{\cal D}_{\ell j}],
\end{eqnarray}
\begin{equation}
\label{3.23} -\left(\frac{2}{dp}a
P_{xy}^{(0)}+\zeta^{(0)}\right)\left(p\partial_p+T\partial_T\right){\boldsymbol {\cal E}}_{0}- a
c_y\frac{\partial}{\partial c_x}{\boldsymbol {\cal E}}_{0}-J_0[{\boldsymbol {\cal E}}_{0},f^{(0)}]={\bf E}_{0}.
\end{equation}
Here, ${\bf A}_{0}({\bf c})$, ${\bf B}_{0}({\bf c})$, ${\bf C}_{0}({\bf c})$, ${\sf D}_0({\bf c})$, and ${\bf
E}_{0}({\bf c})$ are defined by Eqs.\ (\ref{b9}), (\ref{b10}), (\ref{b11}), (\ref{b12}), and (\ref{b13}),
respectively. In addition, upon writing Eqs.\ (\ref{3.19})--(\ref{3.23}), use has been made of the explicit form
of $f^{(1)}$. It has been derived in Ref.\ \cite{G06} and reads
\begin{equation}
\label{3.24} f^{(1)}={\boldsymbol {\cal B}}\cdot \nabla p+{\boldsymbol {\cal C}}\cdot \nabla T+{\boldsymbol
{\cal D}}:\nabla \delta {\bf u},
\end{equation}
where the coefficients ${\boldsymbol {\cal B}}$, ${\boldsymbol {\cal C}}$, and ${\boldsymbol {\cal D}}$ are
functions of the peculiar velocity ${\bf c}$ and the hydrodynamic fields. In particular, the first-order
contribution to the cooling rate, $\zeta_{u,ij}$, is given by
\begin{equation}
\label{3.24.1} \zeta_{u,ij}=-\frac{1}{dp}\int d{\bf c}\,mc^2\left(J[f^{(0)},{\cal D}_{ij}]+J[{\cal
D}_{ij},f^{(0)}]\right).
\end{equation}

The first-order contribution to the mass flux ${\bf j}_0^{(1)}$ of the impurity is defined as
\begin{equation}
{\bf j}_{0}^{(1)}=m_{0}\int \dd{\bf c}\,{\bf c}\,f_0^{(1)}({\bf c}).\label{3.25}
\end{equation}
Use of Eq.\ (\ref{3.17}) into Eq.\ (\ref{3.25}) gives the expression
\begin{equation}
\label{3.26} j_{0,i}^{(1)}=-m_0D_{ij}\frac{\partial x_0}{\partial r_j}-\frac{m}{T}D_{p,ij}\frac{\partial
p}{\partial r_j}-\frac{mn}{T}D_{T,ij}\frac{\partial T}{\partial r_j}+\chi_{ij}{\cal F}_j,
\end{equation}
where
\begin{equation}
\label{3.27} D_{ij}=-\int \dd{\bf c}\,c_i\;{\cal A}_{0,j}({\bf c}),
\end{equation}
\begin{equation}
\label{3.28} D_{p,ij}=-\frac{Tm_0}{m}\int \dd{\bf c}\,c_i\;{\cal B}_{0,j}({\bf c}),
\end{equation}
\begin{equation}
\label{3.29} D_{T,ij}=-\frac{Tm_0}{mn}\int \dd{\bf c}\,c_i\;{\cal C}_{0,j}({\bf c}),
\end{equation}
\begin{equation}
\label{3.30} \chi_{ij}=m_{0}\int \dd{\bf c}\,c_i\;{\cal E}_{0,j}({\bf c}).
\end{equation}
Upon writing Eqs.\ (\ref{3.27})--(\ref{3.30}) use has been made of the symmetry properties of ${\boldsymbol
{\cal A}}_{0}$, ${\boldsymbol {\cal B}}_{0}$, ${\boldsymbol {\cal C}}_{0}$,
 and ${\boldsymbol {\cal E}}_{0}$. In general,
the set of {\em generalized} transport coefficients $D_{ij}$, $D_{p,ij}$, $D_{T,ij}$, and $\chi_{ij}$ are
nonlinear functions of the shear rate and the coefficients of restitution $\alpha$ and $\alpha_0$. It is
apparent that the anisotropy induced by the presence of shear flow gives rise to new transport coefficients for
the mass flux, reflecting broken symmetry. According to Eq.\ (\ref{3.26}), the mass flux of the impurity is
expressed in terms of a diffusion tensor $D_{ij}$, a pressure diffusion tensor $D_{p,ij}$, a thermal diffusion
tensor $D_{T,ij}$, and a mobility tensor $\chi_{ij}$.  Note that in the particular case of the gravitational
force ${\bf F}=m{\bf g}$ and ${\bf F}_0=m_0{\bf g}$, where ${\bf g}$ is the gravity acceleration. In this case,
the combined force ${\boldsymbol {\cal F}}$ defined in Eq.\ (\ref{3.18}) vanishes. Consequently, the external
force ${\boldsymbol {\cal F}}$ does not occur in Eq.\ (\ref{3.17}) when the system is only subjected to a
gravity field.

\subsection{Steady state conditions}

The evaluation of the above transport coefficients requires to know the complete dependence of $P_{ij}^{(0)}$
and $P_{0,ij}^{(0)}$ on the pressure $p$ and the temperature $T$. This involves the corresponding numerical
integrations of the differential equations obeying the pressure tensors $P_{ij}^{(0)}$ and $P_{0,ij}^{(0)}$ [see
Eq.\ (\ref{3.15}) for $P_{0,ij}^{(0)}$]. Needless to say, this is quite an intricate problem. However, some
simplifications occur if attention is restricted to linear deviations from the USF steady state described in
Sec.\ \ref{sec2}. In particular, since the contributions to the mass flux (\ref{3.26}) are already of first
order in the deviations from the steady state, one only needs to know the transport coefficients to zero order
in the deviations. This means that $\partial_t^{(0)}T=\partial_t^{(0)}p=0$ and so the term
$(2/dp)aP_{xy}^{(0)}+\zeta^{(0)}=0$ in the left hand side of Eqs.\ (\ref{3.19})--(\ref{3.23}). The differential
equations for the transport coefficients thus become simple coupled algebraic equations.

The dependence of $P_{ij}^{(0)}$ on the pressure $p$ and temperature $T$ occurs explicitly and through its
dependence on the reduced shear rate $a^*\propto \sqrt{T}/p$. Consequently,
\begin{equation}
\label{3.32} p\partial_p P_{ij}^{(0)}=p\partial_p p P_{ij}^*(a^*)=p\left(1-a^*\frac{\partial} {\partial
a^*}\right)P_{ij}^*(a^*),
\end{equation}
\begin{equation}
\label{3.33} T\partial_T P_{ij}^{(0)}=T\partial_T p P_{ij}^*(a^*)=\frac{1}{2}p a^*\frac{\partial} {\partial
a^*}P_{ij}^*(a^*),
\end{equation}
where $P_{ij}^*=P_{ij}^{(0)}/p$ and $a^*=a/\nu$. The dependence of $P_{ij}^*$ on $a^*$ near the steady state was
determined in the one-component problem \cite{G06} so that all the terms appearing in the integral equations are
explicitly known in the steady state. Under the above conditions, Eqs.\ (\ref{3.19})--(\ref{3.23}) become
\begin{equation}
\label{3.34} - a c_y\frac{\partial}{\partial c_x}{\boldsymbol {\cal A}}_{0}-J_0[{\boldsymbol {\cal
A}}_{0},f^{(0)}]={\bf A}_{0},
\end{equation}
\begin{eqnarray}
\label{3.35} & & - \left[\frac{2a}{d }(1-a^*\partial_{a^*})P_{xy}^{*}+2\zeta^{(0)}+ a
c_y\frac{\partial}{\partial c_x}\right]{\boldsymbol {\cal B}}_{0}-J_0[{\boldsymbol {\cal B}}_{0},f^{(0)}] ={\bf
B}_{0}\nonumber\\
& & -\left(\frac{2aT}{d p}a^*\partial_{a^*}P_{xy}^{*}-\frac{T\zeta^{(0)}}{p} \right){\boldsymbol {\cal
C}}_{0}+J_0[f_0^{(0)},{\boldsymbol {\cal B}}],
\end{eqnarray}
\begin{eqnarray}
\label{3.36}  & & -\left[\frac{2a}{d }(1+\frac{1}{2}a^*\partial_{a^*})P_{xy}^{*}+\frac{1}{2}\zeta^{(0)}+ a
c_y\frac{\partial}{\partial c_x}\right] {\boldsymbol {\cal C}}_{0}-J_0[{\boldsymbol {\cal C}}_{0},f^{(0)}]={\bf
C}_{0}\nonumber\\
&& +\frac{p}{T}\left(\frac{a}{d}a^*\partial_{a^*}P_{xy}^{*}- \frac{\zeta^{(0)}}{2} \right){\boldsymbol {\cal
B}}_{0}+J_0[f_0^{(0)},{\boldsymbol {\cal C}}],
\end{eqnarray}
\begin{equation}
\label{3.37}  - a c_y\frac{\partial}{\partial c_x}{\cal D}_{0,\ell j}-a\delta_{\ell y}{\cal D}_{0,xj}-J_0[{\cal
D}_{0,\ell j},f^{(0)}]=D_{0,\ell j}+\zeta_{u,\ell j}
\left(p\partial_p+T\partial_T\right)f^{(0)}+J_0[f_0^{(0)},{\cal D}_{\ell j}],
\end{equation}
\begin{equation}
\label{3.38} - a c_y\frac{\partial}{\partial c_x}{\boldsymbol {\cal E}}_{0}-J_0[{\boldsymbol {\cal
E}}_{0},f^{(0)}]={\bf E}_{0}.
\end{equation}
It must be recalled that in Eqs.\ (\ref{3.34})--(\ref{3.38}) all the quantities are evaluated in the steady
state, namely, $P_{ij}^*$ and $a^*$ are given by Eqs.\ (\ref{a1})--(\ref{a3}) and (\ref{a4}), respectively,
while $\partial P_{ij}^*/\partial a^*$ is given by Eqs.\ (\ref{c1})--(\ref{c3}). Henceforth, the calculations
will be restricted to the particular condition (\ref{2.19}).

\section{Mass transport of the impurity\label{sec4}}

This Section is devoted to the determination of the generalized transport coefficients associated with the mass
transport of the impurity. In order to get explicit expressions for these coefficients, one has to know the
quantities ${\boldsymbol {\cal A}}_0$, ${\boldsymbol {\cal B}}_0$, ${\boldsymbol {\cal C}}_0$, and ${\boldsymbol
{\cal E}}_0$ which verify the coupled integral equations (\ref{3.34}), (\ref{3.35}), (\ref{3.36}), and
(\ref{3.38}), respectively. To determine the explicit dependence of these quantities on the coefficients of
restitution $\alpha$ and $\alpha_0$, one needs to make use of certain approximations. The standard approach is
to consider the leading term in a Sonine polynomial expansion. In a previous work on diffusion in shear flow
\cite{G02}, the isotropic part of $f_0^{(1)}$ in this expansion was assumed  for simplicity to be a Maxwellian
$f_{0,M}({\bf c})$. However, given that the system is strongly sheared, it is reasonable to expect that the
isotropic part of $f_0^{(1)}$ is mainly governed by the shear flow distribution $f_{0}^{(0)}$ rather than by the
Maxwellian distribution. For this reason, here we keep the usual structure of the standard Sonine approximation,
except that the Maxwellian weight function $f_{0,M}$ is replaced by $f_{0}^{(0)}$. A similar type of modified
Sonine approximation has been recently considered \cite{GSM06} to estimate the Navier-Stokes transport
coefficients of a single granular gas.

According to the above arguments, in the case of the mass flux ${\bf j}_0^{(1)}$, a good estimate of
$\{{\boldsymbol {\cal A}}_0, {\boldsymbol {\cal B}}_0, {\boldsymbol {\cal C}}_0, {\boldsymbol {\cal E}}_0 \}$ is
given by the first Sonine approximation:
\begin{subequations}
\begin{equation}
\label{4.1} {\cal A}_{0,i}\to -c_j\overline{D}_{ji}\; f_{0}^{(0)}({\bf c}), \quad {\cal B}_{0,i}\to
-c_j\overline{D}_{p,ji}\; f_{0}^{(0)}({\bf c}),
\end{equation}
\begin{equation}
\label{4.2} {\cal C}_{0,i}\to -c_j\overline{D}_{T,ji}\; f_{0}^{(0)}({\bf c}), \quad {\cal E}_{0,i}\to
c_j\overline{\chi}_{ji}\; f_{0}^{(0)}({\bf c}),
\end{equation}
\end{subequations}
where $f_{0}^{(0)}({\bf c})$ is the solution of Eq.\ (\ref{3.13}). The relationship between the tensors
$\{\overline{D}_{ij}, \overline{D}_{p,ij}, \overline{D}_{T,ij}, \overline{\chi}_{ij}\}$ and their corresponding
counterparts $\{D_{ij}, D_{p,ij}, D_{T,ij}, \chi_{ij}\}$ is consistently obtained from Eqs.\
(\ref{3.27})--(\ref{3.30}). A simple calculation yields
\begin{equation}
\label{4.2.1} \overline{D}_{ij}=m_0Q_{ik}D_{kj},\quad \overline{D}_{p,ij}=\frac{m}{T}Q_{ik}D_{p,kj},
\end{equation}
\begin{equation}
\label{4.2.2} \overline{D}_{T,ij}=\frac{mn}{T}Q_{ik}D_{T,kj},\quad \overline{\chi}_{ij}=Q_{ik}\chi_{kj},
\end{equation}
where ${\sf Q}={\sf P}_0^{-1}$. When one takes $f_{0,M}$ instead of $f_{0}^{(0)}$ in the Sonine expansion, then
$Q_{ij}=(n_0T_0)^{-1}\delta_{ij}$ and one recovers previous results \cite{G02}. Consistently, the quantities
${\boldsymbol {\cal B}}$ and ${\boldsymbol {\cal C}}$ corresponding to the distribution $f^{(1)}$ of the
granular gas [see Eq.\ (\ref{3.24})] must be similarly approximated. However, as shown in Ref.\ \cite{G06},
these quantities vanish in the first Sonine approximation and so there is no contribution to the mass flux
coming from the terms of the form $J_0[f_0^{(0)},{\boldsymbol {\cal B}}]$ and $J_0[f_0^{(0)},{\boldsymbol {\cal
C}}]$. Substitution of the expressions (\ref{4.1}) and (\ref{4.2}) into Eqs.\ (\ref{3.34}), (\ref{3.35}),
(\ref{3.36}), and (\ref{3.38}) gives a closed set of integral equations for $D_{ij}$, $D_{p,ij}$, $D_{T,ij}$,
and $\chi_{ij}$. Multiplication of these equations by $m_0 c_i$ and integration over ${\bf c}$ yields
\begin{equation}
\label{4.4} \left(a_{ik}+\Omega_{ik}\right)D_{kj}=\frac{p}{m_0}P_{0,ij}^*,
\end{equation}
\begin{eqnarray}
\label{4.5}  \left[\frac{2a}{d }(1-a^*\partial_{a^*})P_{xy}^{*}+2\zeta^{(0)}\right]D_{p,ij}-&&
\left(a_{ik}+\Omega_{ik}\right)D_{p,kj}=\nonumber\\
& &  \frac{Tx_0}{m}(1-a^*\partial_{a^*})
\left(\frac{m_0}{m}P_{ij}^*-P_{0,ij}^*\right)\nonumber\\
& & +\left(\frac{2a}{d}a^*\partial_{a^*}P_{xy}^{*}-\zeta^{(0)} \right)D_{T,ij},
\end{eqnarray}
\begin{eqnarray}
\label{4.6}  \left[\frac{2a}{d
}(1+\frac{1}{2}a^*\partial_{a^*})P_{xy}^{*}+\frac{1}{2}\zeta^{(0)}\right]D_{T,ij}-&&
\left(a_{ik}+\Omega_{ik}\right)D_{T,kj}=\nonumber\\
& & \frac{1}{2}\frac{Tx_0}{m}a^*\partial_{a^*}
\left(\frac{m_0}{m}P_{ij}^*-P_{0,ij}^*\right)\nonumber\\
& & -\left(\frac{a}{d}a^*\partial_{a^*}P_{xy}^{*}-\frac{1}{2}\zeta^{(0)} \right)D_{p,ij},
\end{eqnarray}
\begin{equation}
\label{4.7} \left(a_{ik}+\Omega_{ik}\right) \chi_{kj}=n_0\delta_{ij}.
\end{equation}
In the above equations, $P_{0,ij}^*=P_{0,ij}^{(0)}/x_0p$ and $\Omega_{ij}=n_0T_0 \Lambda_{ik}Q_{kj}$, where
\begin{equation}
\label{4.8} \Lambda_{ij}=-\frac{m_0}{n_0T_0}\int\,d{\bf c}\,c_i\,J_0[c_jf_{0}^{(0)},f^{(0)}].
\end{equation}
This quantity can be evaluated by using standard integration techniques with the result
\begin{eqnarray}
\label{4.9} \Lambda_{ij}&=&\frac{\sqrt{2}}{4d}\left( \frac{\overline{\sigma}}{\sigma}\right)^{d-1}\nu \mu
(1+\alpha_{0})\left[(1+\theta)\theta\right]^{-1/2} \left\{(d+2)(1+\theta)\delta_{ij}\right.
\nonumber\\
& &\left. +\theta\left(P_{ij}^*-\delta_{ij}\right)+\left[d+3+(d+2)\theta\right]
\left(\gamma^{-1}P_{0,ij}^*-\delta_{ij}\right)\right\},
\end{eqnarray}
where $\nu$ is defined by Eq.\ (\ref{2.19.1}), $\mu=m/(m+m_0)$, $\gamma=T_0/T$ is the temperature ratio and
$\theta=m_0T/mT_0$ is the mean-square velocity of the gas particles relative to that of the impurity. Upon
deriving Eq.\ (\ref{4.9}), use has been made of the Sonine approximations (\ref{a0}) and (\ref{c4.2}) for the
distributions $f^{(0)}$ and $f_0^{(0)}$, respectively, and the nonlinear term proportional to
$\left(P_{ij}^*-\delta_{ij}\right)\left(\gamma^{-1}P_{0,ij}^*-\delta_{ij}\right)$ has been neglected.
\begin{figure}
\includegraphics[width=0.5 \columnwidth,angle=0]{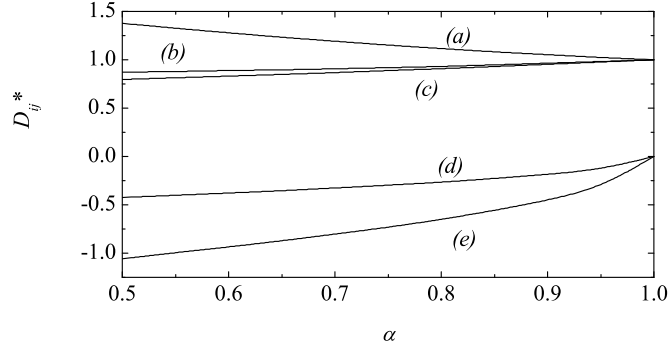}
\caption{Plot of the reduced elements (a) $D_{xx}^*$, (b) $D_{zz}^*$, (c) $D_{yy}^*$, (d) $D_{yx}^*$, and (e)
$D_{xy}^*$ as functions of the (common) coefficient of restitution $\alpha=\alpha_0$ for a three-dimensional
system in the case $\sigma=\overline{\sigma}$ and $m_0/m=0.5$. \label{fig1}}
\end{figure}
\begin{figure}
\includegraphics[width=0.5 \columnwidth,angle=0]{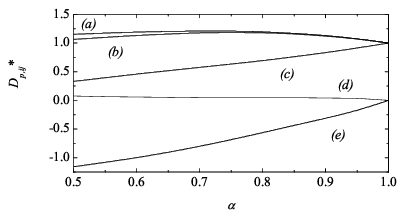}
\caption{Plot of the reduced elements (a) $D_{p,zz}^*$, (b) $D_{p,yy}^*$, (c) $D_{p,xx}^*$, (d) $D_{p,yx}^*$,
and (e) $D_{p,xy}^*$ as functions of the (common) coefficient of restitution $\alpha=\alpha_0$ for a
three-dimensional system in the case $\sigma=\overline{\sigma}$ and $m_0/m=0.5$. \label{fig2}}
\end{figure}
\begin{figure}
\includegraphics[width=0.5 \columnwidth,angle=0]{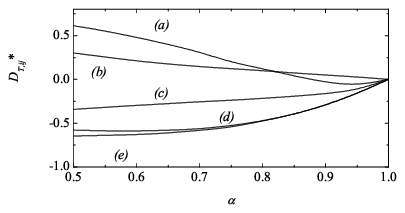}
\caption{Plot of the reduced elements (a) $D_{T,xy}^*$, (b) $D_{T,xx}^*$, (c) $D_{T,yx}^*$, (d) $D_{T,yy}^*$,
and (e) $D_{T,zz}^*$ as functions of the (common) coefficient of restitution $\alpha=\alpha_0$ for a
three-dimensional system in the case $\sigma=\overline{\sigma}$ and $m_0/m=0.5$. \label{fig3}}
\end{figure}
\begin{figure}
\includegraphics[width=0.5 \columnwidth,angle=0]{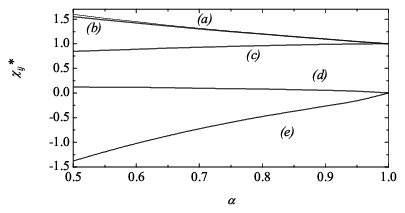}
\caption{Plot of the reduced elements (a) $\chi_{zz}^*$, (b) $\chi_{yy}^*$, (c) $\chi_{xx}^*$, (d) $\chi_{yx}^*$
, and (e) $\chi_{xy}^*$ as functions of the (common) coefficient of restitution $\alpha=\alpha_0$ for a
three-dimensional system in the case $\sigma=\overline{\sigma}$ and $m_0/m=0.5$. \label{fig4}}
\end{figure}

The coefficients $D_{ij}$ and $\chi_{ij}$ decouple from the other ones and hence can be obtained more easily. By
using matrix notation, they are given by
\begin{equation}
\label{4.10} {\sf D}=\frac{p}{m_0}\left({\sf a}+{\boldsymbol {\Omega}}\right)^{-1}\cdot {\sf P}_0^{*},
\end{equation}
\begin{equation}
\label{4.11} {\boldsymbol {\chi}}=n_0\left({\sf a}+{\boldsymbol {\Omega}}\right)^{-1}\cdot {\sf I}.
\end{equation}
The remaining coefficients $D_{p,ij}$ and $D_{T,ij}$ are coupled and they obey the set of simple algebraic
equations (\ref{4.5}) and (\ref{4.6}).

In the elastic limit ($\alpha=\alpha_0=1$, which implies $a^*=0$ in the steady state conditions), $T=T_0$,
$P_{ij}^*=P_{0,ij}^*=\delta_{ij}$, and $\Omega_{ij}=\Omega_0 \delta_{ij}$, so that Eqs.\
(\ref{4.4})--(\ref{4.7}) have the solutions $D_{ij}=D_0\delta_{ij}$, $D_{p,ij}=D_{p0}\delta_{ij}$, $D_{T,ij}=0$,
and $\chi_{ij}=\chi_0\delta_{ij}$, where $D_0$, $D_{p0}$ and $\chi_0$ are the conventional Navier-Stokes
transport coefficients for ordinary gases \cite{CC70}. Their expressions are
\begin{equation}
\label{4.12} D_0=\frac{p}{m_0\Omega_0},\quad D_{p0}=\frac{Tx_0}{m\Omega_0}\left(1-\frac{m_0}{m}\right), \quad
\chi_0=\frac{n_0}{\Omega_0},
\end{equation}
where
\begin{equation}
\label{4.13} \Omega_0=\frac{4}{d}\frac{\pi^{(d-1)/2}}{\Gamma(d/2)}
n\overline{\sigma}^{d-1}\sqrt{\frac{2Tm}{m_0(m+m_0)}}.
\end{equation}
\begin{figure}
\includegraphics[width=0.5 \columnwidth,angle=0]{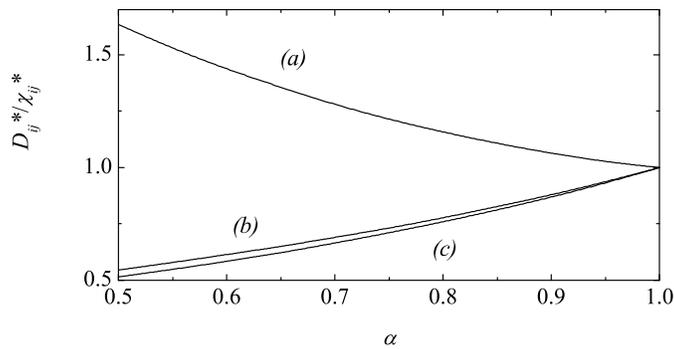}
\caption{Plot of the ratios (a) $D_{xx}^*/\chi_{xx}^*$, (b) $D_{zz}^*/\chi_{zz}^*$, and (c)
$D_{yy}^*/\chi_{yy}^*$ as functions of the (common) coefficient of restitution $\alpha=\alpha_0$ for a
three-dimensional system in the case $\sigma=\overline{\sigma}$ and $m_0/m=0.5$. \label{fig5}}
\end{figure}
To illustrate the dependence of the tensors $T_{ij}\equiv \{D_{ij}, D_{p,ij}, D_{T,ij}, \chi_{ij}\}$ on
dissipation, let us consider a three-dimensional system. In this case, according to Eqs.\
(\ref{4.4})--(\ref{4.7}), $T_{xz}=T_{zx}=T_{yz}=T_{zy}=0$, in agreement with the symmetry of the problem. As a
consequence, there are five relevant elements: the three diagonal ($T_{xx}$, $T_{yy}$, and $T_{zz}$) and two
off-diagonal elements ($T_{xy}$, and $T_{yx}$). The integral equations (\ref{4.4})--(\ref{4.7}) also show that
$T_{xx}\neq T_{yy}\neq T_{zz}$ and $T_{xy}\neq T_{yx}$. In Figs.\ \ref{fig1}--\ref{fig4}, we plot the relevant
reduced elements of tensors $D_{ij}^*$, $D_{p,ij}^*$, $D_{T,ij}^*$ and $\chi_{ij}^*$ as functions of the
(common) coefficient of restitution $\alpha=\alpha_0$ when $\sigma=\overline{\sigma}$ and $m_0/m=0.5$. Here, the
tensors have been reduced with respect to their values in the elastic case, namely, $D_{ij}^*=D_{ij}/D_0$,
$D_{p,ij}^*=D_{p,ij}/D_{p0}$, and $\chi_{ij}^*=\chi_{ij}/\chi_0$, while $D_{T,ij}^*=D_{T,ij}/(x_0T/m\nu)$. We
observe that in general the influence of dissipation on the transport coefficients is quite significant. This
means that the deviation of the elements $T_{ij}$ from their functional forms for elastic collisions is
important for moderate dissipation. It is also apparent that the anisotropy of the system, as measured by the
differences $|T_{xx}-T_{yy}|$ and $|T_{yy}-T_{zz}|$, grows with the inelasticity. This anisotropy is much more
important in the plane of shear flow ($|T_{xx}-T_{yy}|$) than in the plane perpendicular to the flow velocity
($|T_{yy}-T_{zz}|$).

As expected, the usual Einstein relation between the diffusion and mobility coefficients for ordinary fluids
\cite{M89} is no longer valid in this nonequilibrium situation for granular gases. A similar conclusion has been
found when the gas is under HCS \cite{DG01,BLP04,G04}. There are basically two independent reasons for this
violation: the occurrence of different kinetic temperatures between the impurity and gas particles and the
inherent non-Newtonian properties of the reference state. The deviation of the ratio $D_{ij}^*/\chi_{ij}^*$ from
unity is a measure of the violation of Einstein's relation in the USF state. This ratio is plotted in Fig.\
\ref{fig5} for the diagonal elements in the case $\sigma=\overline{\sigma}$ and $m_0/m=0.5$. As Fig.\ \ref{fig5}
clearly shows, the ratio $D_{ij}^*/\chi_{ij}^*\neq 1$ and quickly decays in this situation (when the impurity is
lighter than the gas particles) for the $yy$ and $zz$ elements, while the opposite happens for the $xx$
elements. The violation of the Einstein relation obtained here contrasts with some numerical experiments
performed by Makse and Kurchan \cite{MK02} who applied uniform shear to measure diffusivity and mobility in
bidisperse mixtures. Their results show that Einstein's relation is verified so that a temperature can be
defined in the system by analogy with ordinary fluids. However, this conclusion also disagrees with the lack of
equipartition in a granular sheared gas when different particles are present \cite{CH02}.

\section{Concluding remarks}
\label{sec5}

In this paper, mass transport of an impurity or intruder immersed in a strongly sheared granular gas at
low-density has been analyzed. We have been interested in a situation where {\em weak} spatial gradients of
concentration, temperature and pressure coexist with a {\em strong} shear rate. In addition, we have also
assumed that the system (gas plus impurity) feels the action of external forces which are considered to be at
least of first order in the spatial gradients. Under these conditions, the resulting mass transport process is
anisotropic and thus it cannot be described by scalar transport coefficients. Instead, it must be described by
shear-rate dependent tensorial quantities whose explicit determination has been the main objective of this
paper.

In the tracer limit, the inelastic Boltzmann and Boltzmann-Lorentz kinetic equations describe the state of the
gas and the impurity, respectively. Since the state of the gas slightly deviates from the USF by small spatial
gradients, a generalized Chapman-Enskog method has been recently proposed \cite{L06,G06} to analyze transport
around nonequilibrium states. Here, a similar perturbative scheme has been used to solve the Boltzmann-Lorentz
equation to first order in the deviations of the hydrodynamic field gradients from their values in the reference
shear flow state $f_0^{(0)}$. As noted in previous works  \cite{L06,G06}, the zeroth-order distribution
$f_0^{(0)}$ is not in general stationary and only in very special conditions has a simple relation with the
(steady) USF distribution. Since we are mainly interested in determining mass transport of the impurity ${\bf
j}_0^{(1)}$ in the USF state, for practical purposes the results have been specialized to the steady state
conditions, namely, when the hydrodynamic variables satisfy the relation (\ref{2.19}). This implies that the
reduced shear rate $a^*$ is coupled with the coefficient of restitution $\alpha$, so that the latter is the
relevant parameter of the problem. Under these conditions, ${\bf j}_0^{(1)}$ is given by Eq.\ (\ref{3.26}) where
the corresponding set of generalized transport coefficients $\{D_{ij}, D_{p,ij}, D_{T,ij}, \chi_{ij}\}$ are the
solutions of the linear integral equations (\ref{3.34})--(\ref{3.38}). To get explicit results,  a first Sonine
polynomial approximation has been considered to estimate some collisional integrals. The reliability of this
approach has been assessed in the USF problem, where it has been shown to agree very well with Monte Carlo
simulations \cite{MG02,AS05}.

\begin{figure}
\includegraphics[width=0.5 \columnwidth,angle=0]{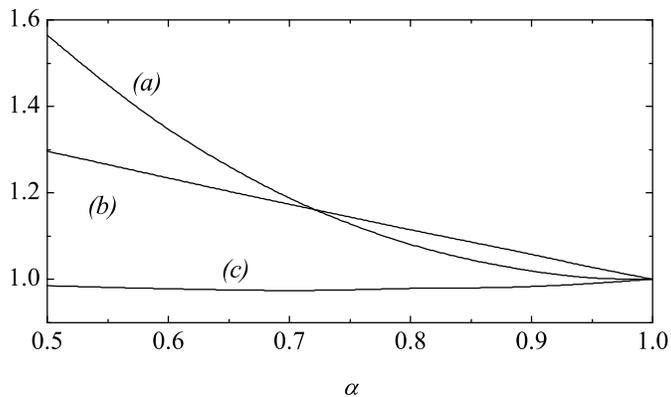}
\caption{Plot of the ratios (a) $D_p/(\frac{1}{d}D_{p,kk})$, (b) $D/(\frac{1}{d}D_{kk})$, and (c)
$\chi/(\frac{1}{d}\chi_{kk})$ as functions of the (common) coefficient of restitution $\alpha=\alpha_0$ for a
three-dimensional system in the case $\sigma=\overline{\sigma}$ and $m_0/m=0.5$. \label{fig6}}
\end{figure}

The results show that the coefficients $\{D_{ij}, D_{p,ij}, D_{T,ij}, \chi_{ij}\}$ present a complex dependence
on the coefficients of restitution $\alpha$ and $\alpha_0$ and on the masses and sizes of the system. This is
clearly illustrated in Figs.\ \ref{fig1}--\ref{fig4}. The deviations of $\{D_{ij}, D_{p,ij}, D_{T,ij},
\chi_{ij}\}$ from their elastic counterparts are basically due to three different reasons. First, the presence
of shear flow modifies the collision frequency of the elastic diffusion problem $\Omega_0$ [defined by Eq.\
(\ref{4.13})] by the tensorial term $a_{ij}+\Omega_{ij}$. Second, given that in general the impurity and gas
particles are mechanically different, the reference shear flow state of the impurity $f_0^{(0)}$ is completely
different from that of the gas particles $f^{(0)}$. This effect gives rise to terms proportional to
$(m_0/m)P_{ij}^*-P_{0,ij}^*$. Third, there is a  coupling between the coefficients $D_{p,ij}$ and $D_{T,ij}$ due
to the inherent non-Newtonian features of the USF state of the gas. Each one of the three effects is a different
reflection of the dissipation present in the system.

Most of the works \cite{mixtures} on granular mixtures have been based on the CE expansion around an elastic
(local) equilibrium state up to the Navier-Stokes order, and therefore they are limited to nearly-elastic
systems in the USF. A more recent CE expansion \cite{GD02,GM06} around the (local) HCS takes into account energy
nonequipartition and provides expressions for the Navier-Stokes transport coefficients of the mixture without
any restriction on the level of inelasticity. In particular, the mass transport of impurity is characterized by
the single scalar coefficients $D$, $D_p$, $D_T$, and $\chi$. Although the base state considered in Refs.\
\cite{GD02} and \cite{GM06} is different from the one chosen here, it is still worthwhile to carry out some
comparison between both descriptions. To that end, let us define the scalars $\frac{1}{d}D_{kk}$,
$\frac{1}{d}D_{p,kk}$, and $\frac{1}{d}\chi_{kk}$. These coefficients can be understood as the generalized
diffusion coefficient, pressure diffusion coefficient, and mobility coefficient in a strongly sheared mixture.
In Fig.\ \ref{fig6}, the ratios $D/(\frac{1}{d}D_{kk})$, $D_p/(\frac{1}{d}D_{p,kk})$, and
$\chi/(\frac{1}{d}\chi_{kk})$ are plotted versus the (common) coefficient of restitution $\alpha=\alpha_0$ for a
three-dimensional system in the case $\sigma=\overline{\sigma}$ and $m_0/m=0.5$. Here, $D$, $D_p$ and $\chi$
refer to the coefficients obtained in Refs.\ \cite{GD02} and \cite{DG01}. As expected, we observe that in
general the transport coefficients of the {\em perturbed} USF state differ from the usual Navier-Stokes
coefficients as the collisions become more inelastic. These discrepancies are specially significant in the cases
of the diffusion and the pressure diffusion coefficients.

It is apparent that the results presented here are relevant to make a comparison with numerical simulations. In
the self-diffusion problem (when the impurity and gas particles are mechanically equivalent), previous results
obtained for the diffusion tensor \cite{G02} have shown good qualitative agreement with molecular dynamics
simulations \cite{C97}. Beyond this particular case, to my knowledge no previous studies on the tensors
$\{D_{ij}, D_{p,ij}, D_{T,ij}, \chi_{ij}\}$ have been performed. I hope that this paper stimulates the
performance of such computer studies to check the relevance of kinetic theory to describe mass transport under
shear flow.

A possible application of the results reported in this paper is to study segregation induced by a thermal
gradient. Thermal diffusion is caused by the relative motion of the components of a mixture due to the presence
of a temperature gradient. Due to this motion, concentration gradients subsequently appear in the mixture
producing diffusion that tends to oppose those gradients. In the steady state, ${\bf j}_0^{(1)}={\bf 0}$ and the
thermal diffusion factor provides a segregation criterion. Recent kinetic theory results \cite{G06bis} based on
the Navier-Stokes transport coefficients have been able to explain some experimental and/or molecular dynamics
segregation results obtained in agitated granular mixtures at large shaking amplitudes \cite{SUKSS06}. Another
possible direction of study is the extension of the present approach to mixtures with finite composition. Given
the mathematical difficulties associated with the description of multicomponent systems in far from equilibrium
situations, one could perhaps use a kinetic model of the Boltzmann equation. Once the transport coefficients of
the mixture are known, a linear stability analysis of the hydrodynamic equations could be carried out to
identify the conditions for instabilities at long wavelengths \cite{G06}. Finally, it must be noted that the
results reported here have been made in the context of a very simple collision model where the coefficients of
restitution $\alpha$ and $\alpha_0$ are constant. However, experiments and simulations show that the
coefficients of restitution depend in general on the relative velocity of colliding particles \cite{BP04}.
Recent results \cite{BP03,BSSP04} derived for these viscoelastic models in the case of the Navier-Stokes
transport coefficients show qualitative differences from the ones obtained with the simplifying assumption of a
constant coefficient of restitution. For this reason, it would be interesting to extend the present description
to this kind of models (where collisions are described by an impact velocity dependent coefficient of
restitution) in order to check whether the behavior predicted here for the generalized transport coefficients
also occurs (at least from a qualitative level) in the case of granular gases of viscoelastic particles. As in
the Navier-Stokes description \cite{BP03}, explicit results for this more realistic model could be obtained only
for small enough dissipation. This contrasts with the results reported here since they are not restricted to any
level of inelasticity. Work along the above lines will be done in the near future.

\acknowledgments

Partial support from the Ministerio de Ciencia y Tecnolog\'{\i}a (Spain) through Grant No. BFM2001-0718 is
acknowledged.

\appendix
\section{Rheological properties in the steady USF
\label{appA}}

The explicit expressions for the pressure tensors $P_{ij}^*\equiv P_{ij}/p$ and $P_{0,ij}^*\equiv P_{0,ij}/x_0p$
of the gas and the impurity in the steady USF are provided in this Appendix. To get the explicit expressions of
the elements of $P_{ij}^*$, one takes the following Sonine approximation for $f$:
\begin{equation}
\label{a0} f({\bf V})\to f_{M}({\bf V}) \left[1+\frac{m}{2T}\left(\frac{P_{k\ell}}{nT}
-\delta_{k\ell}\right)\left(V_{k}V_{\ell} -\frac{1}{d}V^2\delta_{k\ell}\right)\right] ,
\end{equation}
where
\begin{equation}
\label{a0.1} f_{M}({\bf V})=n \left(\frac{m}{2\pi T}\right)^{d/2}\exp\left(-\frac{mV^2}{2T}\right).
\end{equation}
By using this approximation, the nonzero elements of $P_{ij}^*$ are given by \cite{SGD04}
\begin{equation}
\label{a1} P_{yy}^*=P_{zz}^*=\cdots=P_{dd}^*=\frac{d+1+(d-1)\alpha} {2d+3-3\alpha}\;,
\end{equation}
\begin{equation}
\label{a2} P_{xy}^*=-4d\frac{d+1+(d-1)\alpha} {(1+\alpha)(2d+3-3\alpha)^2}a^*\;,
\end{equation}
\begin{equation}
\label{a3} P_{xx}^*=d-(d-1)P_{yy}^*.
\end{equation}
The relationship between the reduced shear rate $a^*=a/\nu$ [where $\nu$ is defined by Eq.\ (\ref{2.19.1})] and
the coefficient of restitution $\alpha$ is
\begin{equation}
\label{a4} a^{*2}=\frac{d+2}{32d} \frac{(1+\alpha)(2d+3-3\alpha)^2(1-\alpha^2)}{ d+1+(d-1)\alpha}\;.
\end{equation}
Moreover, the (reduced) cooling rate $\zeta^*=\zeta^{(0)}/\nu$ is
\begin{equation}
\label{a5} \zeta^*=\frac{d+2}{4d}(1-\alpha^2).
\end{equation}

In the case of $P_{0,ij}^*$, one considers the leading Sonine approximation for $f_0^{(0)}$ given by
\begin{equation}
\label{c4.2} f_0({\bf V})\to f_{0,M}({\bf V}) \left[1+\frac{m_0}{2T_0}\left(\frac{P_{0,k\ell}}{n_0T_0}
-\delta_{k\ell}\right)\left(V_{k}V_{\ell} -\frac{1}{d}V^2\delta_{k\ell}\right)\right] ,
\end{equation}
where
\begin{equation}
\label{4.3} f_{0,M}({\bf V})=n_0 \left(\frac{m_0}{2\pi T_0}\right)^{d/2}\exp\left(-\frac{m_0V^2}{2T_0}\right).
\end{equation}
Another different approach based on an anisotropic Gaussian distribution has been considered by Lutsko
\cite{L04}. This latter approximation has the additional advantage of being positive definite. However, the
algebraic equations defining the pressure tensor $P_{0,ij}^*$ cannot be solved explicitly since it requires to
numerically solve some collision integrals. Here, for the sake of simplicity, we have preferred to estimate the
rheological properties of the system by means of Grad's solution (\ref{c4.2}). This allows us to get explicit
expressions for $P_{0,ij}^*$. It has been shown, by comparison to DSMC simulations \cite{MG02,L04}, that the
results derived from both approaches compare very well with computer simulations, even for strong dissipation.
Using (\ref{c4.2}), the nonzero elements of $P_{0,ij}^*$ can be written as \cite{G02}
\begin{equation}
\label{a6} P_{0,yy}^*=P_{0,zz}^*=\cdots=P_{0,dd}^*=-\frac{F+HP_{yy}^*}{G},
\end{equation}
\begin{equation}
\label{a7} P_{0,xy}^*=\frac{a^*P_{0,yy}^*-HP_{xy}^*}{G},
\end{equation}
\begin{equation}
\label{a8} P_{0,xx}^*=d\gamma-(d-1)P_{0,yy}^*,
\end{equation}
where $\gamma=T_0/T$ is the temperature ratio and
\begin{equation}
\label{a9} F=\frac{\sqrt{2}}{2d}\left(\frac{\overline{\sigma}}{\sigma}
\right)^{d-1}\mu_{0}\left(\frac{1+\theta}{\theta^3}\right)^{1/2}(1+\alpha_{0})
\left[1+\frac{\mu}{2}(d-1)(1+\theta)(1+\alpha_{0})\right],
\end{equation}
\begin{eqnarray}
\label{a10} G&=&-\frac{\sqrt{2}}{4d}\left(\frac{\overline{\sigma}}{\sigma}
\right)^{d-1}\mu\left(\frac{1}{\theta(1+\theta)}\right)^{1/2}
(1+\alpha_{0})\nonumber\\
& & \times \left\{2[(d+2)\theta+d+3]-3\mu (1+\theta)(1+\alpha_{0})\right\},
\end{eqnarray}
\begin{equation}
\label{a11} H=\frac{\sqrt{2}}{4d}\left(\frac{\overline{\sigma}}{\sigma}
\right)^{d-1}\mu_{0}\left(\frac{1}{\theta(1+\theta)}\right)^{1/2}
(1+\alpha_{0})\left[3\mu(1+\theta)(1+\alpha_{0})-2\right].
\end{equation}
Here, $\theta=m_0/m\gamma$, $\mu=m/(m+m_0)$, and $\mu_0=1-\mu=m_0/(m+m_0)$. The temperature ratio $\gamma$ is
determined from the condition
\begin{equation}
\label{a12} \gamma=\frac{\zeta^*P_{0,xy}^*}{\zeta_0^*P_{xy}^*},
\end{equation}
where the ``cooling rate'' $\zeta_0^*=\zeta_0/\nu$ for the impurity is given by
\begin{equation}
\label{a13} \zeta_0^*=\frac{(d+2)\sqrt{2}}{2d}\left(\frac{\overline{\sigma}}{\sigma}
\right)^{d-1}\mu\left(\frac{1+\theta}{\theta}\right)^{1/2}(1+\alpha_{0})
\left[1-\frac{\mu}{2}(1+\theta)(1+\alpha_{0})\right].
\end{equation}
For elastic collisions ($\alpha=\alpha_0=1$), Eqs.\ (\ref{a1})--(\ref{a3}) and (\ref{a6})--(\ref{a11}) lead to
$P_{ij}^*=P_{0,ij}^*=\delta_{ij}$. In this case, as expected, the solution to Eq.\ (\ref{a12}) is $\gamma=1$. In
addition, if we assume that particles of the gas and the impurity are mechanically equivalent (i.e., $m=m_0$,
$\sigma=\sigma_0$, and $\alpha=\alpha_0$), then $P_{ij}^*=P_{0,ij}^*$, $\zeta^*=\zeta_0^*$ and so $\gamma=1$.
Beyond these two limit cases, the temperatures $T$ and $T_0$ are different so that there is a violation of
energy equipartition.

\section{Behavior of the zeroth-order pressure tensors near the
steady state \label{appC}}

This Appendix addresses the behavior of the pressure tensors $P_{ij}^*$ and $P_{0,ij}^*$ of the gas particles
and the impurity, respectively, near the steady state. The behavior of the second-degree moment $P_{ij}^*$ was
studied in Ref.\ \cite{G06} in the three-dimensional case ($d=3$). The extension to an arbitrary number of
dimensions is straightforward and here only the final expressions are displayed. The behavior of the
$yy$-element is given by
\begin{equation}
\label{c1} \left(\frac{\partial P_{yy}^*}{\partial a^*}\right)_s=4 P_{yy}^*\frac{a_s^*
\Delta+P_{xy}^*}{2a_s^{*2} \Delta +d( 2\beta+\zeta^*)},
\end{equation}
where $a_s^*(\alpha)$ is the steady state value of $a^*(p,T)$ given by Eq.\ (\ref{a4}),
\begin{equation}
\label{c2} \beta=\frac{1+\alpha}{2}\left[1-\frac{d-1}{2d}(1-\alpha)\right],
\end{equation}
and $\Delta\equiv \left(\partial P_{xy}^*/\partial a^*\right)_s$ is the real root of the cubic equation
\begin{equation}
\label{c3} 2 a_s^{*4} \Delta^3+4da_s^{*2}(\zeta^*+\beta)\Delta^2+\frac{d^2}{2}(7\zeta^{*2}
+14\zeta^*\beta+4\beta^2)\Delta+d^2\beta(\zeta^*+\beta)^{-2}( 2\beta^2-2\zeta^{*2}-\beta\zeta^*).
\end{equation}
In the above equations, it is understood that all the quantities are computed in the steady state.

Let us consider now the elements of $P_{0,ij}^*$. In dimensionless form, they verify the equation
\begin{equation}
\label{c4} -\left(\frac{2}{d}a^* P_{xy}^{*}+\zeta^{*}\right)\left(1-\frac{1}{2}a^*\frac{\partial}{\partial
a^*}\right) P_{0,ij}^{*}+ a_{i\ell}^*P_{0,j\ell}^{*}+a_{j\ell}^*P_{0,i\ell}^{*}= B_{ij}^*,
\end{equation}
where
\begin{equation}
\label{c4.1} B_{ij}^*=\frac{m_0}{x_0\nu p} \;\int \dd{\bf c}\,c_ic_j\;J_0[f_0^{(0)},f^{(0)}].
\end{equation}
Upon deriving Eq.\ (\ref{c4}), use has been made of the fact that in the hydrodynamic regime the dimensionless
pressure tensor depends on $p$ and $T$ only through its dependence on the reduced shear rate $a^*=a/\nu(p,T)$
[see Eqs.\ (\ref{3.32}) and (\ref{3.33})]. The collisional moment $B_{ij}^*$ can be estimated by using Grad's
approximation (\ref{c4.2}) with the result \cite{G02}
\begin{equation}
\label{c5} B_{ij}^*=Y \delta_{ij}+X_0 P_{0,ij}^*+X P_{ij}^*,
\end{equation}
where
\begin{equation}
\label{c6} Y=\frac{d+2}{2\sqrt{2}d}\left(\frac{\overline{\sigma}}{\sigma}\right)^{d-1}\mu_0
(1+\alpha_0)\left(\frac{1+\theta}{\theta}\right)^{3/2} \left[\frac{\lambda_{0}}{d+2}+\frac{d}{d+3}\mu
(1+\alpha_{0})\right],
\end{equation}
\begin{equation}
\label{c7} X_0=-\frac{d+2}{\sqrt{2}d}\left(\frac{\overline{\sigma}}{\sigma}\right)^{d-1}
\mu_0(1+\alpha_0)\left[\theta(1+\theta)\right]^{-1/2} \left[
1+\frac{(d+3)}{2(d+2)}\frac{1+\theta}{\theta}\lambda_{0}\right] \gamma^{-1},
\end{equation}
\begin{equation}
\label{c8} X=\frac{d+2}{\sqrt{2}d}\left(\frac{\overline{\sigma}}{\sigma}\right)^{d-1}
\mu_0(1+\alpha_0)\left[\theta(1+\theta)\right]^{-1/2} \left[1-\frac{(d+3)}{2(d+2)}(1+\theta)\lambda_{0}\right],
\end{equation}
with
\begin{equation}
\label{c9} \lambda_{0}=\frac{2}{1+\theta}-\frac{3}{d+3}\mu_0(1+\alpha_{0}).
\end{equation}
Let us consider the elements $P_{0,xy}^*$ and $P_{0,yy}^*$. From Eq.\ (\ref{c4}), one gets
\begin{equation}
\label{c10} -\left(\frac{2}{d}a^* P_{xy}^{*}+\zeta^{*}\right)\left(1-\frac{1}{2}a^*\frac{\partial} {\partial
a^*}\right)P_{0,xy}^* + a^*P_{0,yy}^{*}-X_0P_{0,xy}^*= X P_{xy}^{*},
\end{equation}
\begin{equation}
\label{c11} -\left(\frac{2}{d}a^* P_{xy}^{*}+\zeta^{*}\right)\left(1-\frac{1}{2}a^*\frac{\partial} {\partial
a^*}\right)P_{0,yy}^*-X_0P_{0,yy}^*= Y+X P_{yy}^{*}.
\end{equation}
This set of equations can be written as
\begin{equation}
\label{c12} \frac{\partial P_{0,xy}^*}{\partial a^*}=2\frac{(\zeta^*+\frac{2}{d}a^*P_{xy}^*+X_0)P_{0,xy}^*
-a^*P_{0,yy}^*+XP_{xy}^*}{a^*\left(\zeta^*+\frac{2}{d}a^*P_{xy}^*\right)},
\end{equation}
\begin{equation}
\label{c13} \frac{\partial P_{0,yy}^*}{\partial a^*}=2\frac{(\zeta^*+\frac{2}{d}a^*P_{xy}^*+X_0)P_{0,yy}^*
+Y+XP_{yy}^*}{a^*\left(\zeta^*+\frac{2}{d}a^*P_{xy}^*\right)}.
\end{equation}
It must be remarked that the temperature ratio $\gamma=T_0/T$ is also a function of $a^*$ in the hydrodynamic
solution. Since $P_{0,xx}^*+(d-1)P_{0,yy}^*=d \gamma$, the corresponding equation for the derivative $(\partial
\gamma/\partial a^*)$ can be obtained from Eq.\ (\ref{c4}) as
\begin{equation}
\label{c14} \gamma-\frac{a^*}{2}\frac{\partial \gamma}{\partial
a^*}=\frac{\gamma\zeta_0^*+\frac{2}{d}a^*P_{0,xy}^*} {\zeta^*+\frac{2}{d}a^*P_{xy}^*},
\end{equation}
where $\zeta_0^*$ is given by Eq.\ (\ref{a13}).  The set of three coupled equations (\ref{c12}), (\ref{c13}),
and (\ref{c14}) has a singular point corresponding to the steady state solution, i.e., when
$a^*(p,T)=a_s^*(\alpha)$. In this limit ($a^*\to a_s^*$), the numerators and denominators of Eqs.\ (\ref{c12}),
(\ref{c13}), and (\ref{c14}) vanish. The limit can be evaluated by means of l'Hopital's rule. Thus, when one
differentiates with respect to $a^*$ the numerators and denominators of Eqs.\ (\ref{c12}), (\ref{c13}), and
(\ref{c14}) and then takes the limit $a\to a_s^*$, one gets the relations
\begin{eqnarray}
\label{c15} \left(\frac{\partial P_{0,xy}^*}{\partial
a^*}\right)_s&=&\left(\frac{1}{2}a_s^*\chi_s-X_0\right)^{-1}\left\{ \chi_s
P_{0,xy}^*+\left(X_0'P_{0,xy}^*+X'P_{xy}^*\right)(\partial\gamma/\partial a^*)_s\right.\nonumber\\
& & \left.- \left[P_{0,yy}^*+a_s^*(\partial P_{0,yy}^*/\partial a^*)_s\right]+X(\partial P_{xy}^*/\partial
a^*)_s\right\} ,
\end{eqnarray}
\begin{equation}
\label{c16} \left(\frac{\partial P_{0,yy}^*}{\partial a^*}\right)_s=\frac{\chi_s
P_{0,yy}^*+\left(Y'+X_0'P_{0,yy}^*+X'P_{yy}^*\right)(\partial\gamma/\partial a^*)_s+X (\partial
P_{yy}^*/\partial a^*)_s} {\frac{1}{2}a_s^*\chi_s-X_0},
\end{equation}
\begin{equation}
\label{c17} \left(\frac{\partial \gamma}{\partial a^*}\right)_s=\frac{\chi_s
\gamma-\frac{2}{d}\left[P_{0,xy}^*+a_s^*(\partial P_{0,xy}^*/\partial
a^*)_s\right]}{\frac{1}{2}a_s^*\chi_s+\zeta_0^*+\gamma \zeta_0'},
\end{equation}
where the subscript $s$ means that the derivatives are computed in the steady state. In addition, we have
introduced the quantities
\begin{equation}
\label{c18} \chi_s=\frac{2}{d}\left[P_{xy}^*+a_s^*(\partial P_{xy}^*/\partial a^*)_s\right],
\end{equation}
and $Y'\equiv (\partial Y/\partial \gamma)$, $X'\equiv (\partial X/\partial \gamma)$, $X_0'\equiv (\partial
Y/\partial \gamma)$, and $\zeta_0'\equiv (\partial \zeta_0^*/\partial \gamma)$. Note that all the quantities
appearing on the right-hand side of Eqs.\ (\ref{c15}), (\ref{c16}), and (\ref{c17}) are evaluated in the steady
state. This set of equations can be easily solved to give the corresponding derivatives. The result is
\begin{equation}
\label{c19} \left(\frac{\partial \gamma}{\partial a^*}\right)_s=\frac{\Lambda_1}{\Lambda_2},
\end{equation}
where
\begin{eqnarray}
\label{c20} \Lambda_1&=&d\left(\frac{1}{2}a_s^*\chi_s-X_0\right)\left\{
\left(\frac{1}{2}a_s^*\chi_s-X_0\right)\left(\chi_s\gamma-\frac{2}{d} P_{0,xy}^*\right)\right. \nonumber\\
& & \left. -\frac{2}{d}a_s^*\left[\chi_sP_{0,xy}^*-P_{0,yy}^*+X(\partial P_{xy}^*/\partial a^*)_s\right]\right\}
\nonumber\\
& & +2a_s^{*2}\left[\chi_sP_{0,yy}^*+X(\partial P_{yy}^*/\partial a^*)_s\right],
\end{eqnarray}
\begin{eqnarray}
\label{c21} \Lambda_2&=&d\left(\frac{1}{2}a_s^*\chi_s-X_0\right)\left[
\left(\frac{1}{2}a_s^*\chi_s-X_0\right)\left(\zeta_0^*+\frac{1}{2}\chi_s+\gamma\zeta_0'\right) \right.
\nonumber\\
& &\left.
+\frac{2}{d}a_s^*\left(X_0'P_{0,xy}^*+X'P_{yy}^*\right)\right]-2a_s^{*2}\left(Y'+X_0'P_{0,yy}^*+X'P_{yy}^*\right).
\end{eqnarray}
Substitution of Eq.\ (\ref{c19}) into Eqs.\ (\ref{c15}) and (\ref{c16}) gives the derivatives
$(\partial_{a^*}P_{0,xy}^*)_s$ and $(\partial_{a^*}P_{0,yy}^*)_s$, respectively.

\section{Chapman-Enskog -like expansion
\label{appB}}

In this Appendix, we provide some technical details on the determination of the first-order approximation
$f_0^{(1)}$ by means of the Chapman-Enskog-like expansion. Inserting the expansions (\ref{3.7}) and (\ref{3.9})
into Eq.\ (\ref{3.1}), one gets the kinetic equation for $f_0^{(1)}$
\begin{equation}
\label{b1} \left(\partial_t^{(0)}-aV_y\frac{\partial}{\partial V_x}\right)f_0^{(1)}-J_0[f_0^{(1)},f^{(0)}]=
-\left[\partial_t^{(1)}+({\bf V}+{\bf u}_s)\cdot \nabla +\frac{{\bf F}_0}{m_0}\cdot \frac{\partial}{\partial
{\bf V}} \right]f^{(0)}.
\end{equation}
The velocity dependence on the right side of Eq.\ (\ref{b1}) can be obtained from the macroscopic balance
equations (\ref{3.2})--(\ref{3.3.1}) to first order in the gradients. They are given by
\begin{equation}
\label{b2}
\partial_t^{(1)}x_0=-({\bf u}_s+\delta {\bf u})\cdot \nabla x_0,
\end{equation}
\begin{equation}
\label{b3}
\partial_t^{(1)}\delta {\bf u}=-({\bf u}_s+\delta {\bf u})\cdot \nabla \delta {\bf u}
-\frac{1}{\rho}\left(\nabla \cdot {\sf P}^{(0)}-n{\bf F}\right),
\end{equation}
\begin{equation}
\label{b4} \partial_t^{(1)}p=-({\bf u}_s+ \delta {\bf u})\cdot \nabla p-p(\nabla \cdot \delta {\bf
u}+\zeta^{(1)})-\frac{2}{d}\left( aP_{xy}^{(1)}+{\sf P}^{(0)}:\nabla \delta {\bf u}\right),
\end{equation}
\begin{equation}
\label{b5} \partial_t^{(1)}T=-({\bf u}_s+ \delta {\bf u})\cdot \nabla T-\frac{2}{d n}\left( aP_{xy}^{(1)}+{\sf
P}^{(0)}:\nabla \delta {\bf u}\right)-p\zeta^{(1)},
\end{equation}
where $\rho=mn$ is the mass density,
\begin{equation}
\label{b6} P_{ij}^{(1)}=\int d{\bf c}\, m c_i c_j  f^{(1)}({\bf c}),
\end{equation}
and
\begin{equation}
\label{b7} \zeta^{(1)}=-\frac{1}{dp}\int d{\bf c}\,mc^2\left(J[f^{(0)},f^{(1)}]+J[f^{(1)},f^{(0)}]\right).
\end{equation}
Use of Eqs.\ (\ref{b2})--(\ref{b5}) in Eq.\ (\ref{b1}) yields
\begin{eqnarray}
\label{b8} \left(\partial_t^{(0)}-aV_y\frac{\partial}{\partial V_x}\right)f_0^{(1)}&-&J_0[f_0^{(1)},f^{(0)}]=
{\bf A}_0\cdot \nabla x_0+{\bf B}_0\cdot \nabla p \nonumber\\
& & +{\bf C}_0\cdot \nabla T+{\sf D}_0:\nabla \delta {\bf u} +{\bf E}_0\cdot {\boldsymbol {\cal
F}}\nonumber\\
& & +\zeta^{(1)}(p\partial_p+T\partial_T)f_0^{(0)}+J_0[f_0^{(0)},f^{(1)}],
\end{eqnarray}
where we have introduced the force ${\boldsymbol {\cal F}}={\bf F}_0-(m_0/m){\bf F}$. The coefficients of the
field gradients on the right-hand side of (\ref{b8}) are functions of ${\bf c}$ and the hydrodynamic fields.
They are given by
\begin{equation}
\label{b9} A_{0,i}({\bf c})=-\frac{\partial f_0^{(0)}}{\partial x_0}c_i,
\end{equation}
\begin{equation}
\label{b10} B_{0,i}({\bf c})=-\frac{\partial f_0^{(0)}}{\partial p} c_i+\frac{1}{\rho} \frac{\partial
f_0^{(0)}}{\partial \delta u_j}\frac{\partial P_{ij}^{(0)}}{\partial p},
\end{equation}
\begin{equation}
\label{b11} C_{0,i}({\bf c})=-\frac{\partial f_0^{(0)}}{\partial T} c_i+\frac{1}{\rho} \frac{\partial
f_0^{(0)}}{\partial \delta u_j}\frac{\partial P_{ij}^{(0)}}{\partial T},
\end{equation}
\begin{equation}
\label{b12} D_{0,ij}({\bf c})=p\frac{\partial f_0^{(0)}}{\partial p}\delta_{ij}-\frac{\partial f_0^{(0)}}
{\partial \delta u_i}c_j+\frac{2}{dp}\left(P_{ij}^{(0)}-a\eta_{xyij}\right) \left(p\frac{\partial}{\partial
p}+T\frac{\partial}{\partial T}\right)f_0^{(0)},
\end{equation}
\begin{equation}
\label{b13} E_{0,i}({\bf c})=-\frac{1}{m_0}\frac{\partial f_0^{(0)}}{\partial c_i}.
\end{equation}
Upon writing Eq.\ (\ref{b12}) use has been made of the expression of the pressure tensor $P_{ij}^{(1)}$ of the
gas \cite{G06}
\begin{equation}
\label{b14} P_{ij}^{(1)}=-\eta_{ijk\ell} \frac{\partial \delta u_k} {\partial r_{\ell}},
\end{equation}
where $\eta_{ijk\ell}$ is the viscosity tensor. Moreover, the expression for the distribution $f^{(1)}$ derived
in Ref.\ \cite{G06} is given by Eq.\ (\ref{3.24}).

The solution to Eq.\ (\ref{b8}) has the form
\begin{equation}
 f_0^{(1)}={\boldsymbol {\cal A}}_{0}\cdot \nabla x_0+
{\boldsymbol {\cal B}}_{0}\cdot \nabla p+{\boldsymbol {\cal C}}_{0}\cdot \nabla T+{\sf {\cal D}}_{0}:\nabla
\delta {\bf u}+{\boldsymbol {\cal E}}_{0}\cdot {\boldsymbol {\cal F}}. \label{b15}
\end{equation}
The coefficients ${\boldsymbol {\cal A}}_{0}$, ${\boldsymbol {\cal B}}_{0}$, ${\boldsymbol {\cal C}}_{0}$,
${\boldsymbol {\cal D}}_{0}$, and ${\boldsymbol {\cal E}}_{0}$ are functions of the peculiar velocity and the
hydrodynamic fields $x_0$, $p$, $T$, and $\delta {\bf u}$. The cooling rate $\zeta^{(0)}$ depends on space
through its dependence on $p$ and $T$. Moreover, there are contributions from $\partial_t^{(0)}$ acting on the
pressure, temperature, and velocity gradients given by
\begin{eqnarray}
\label{b16}
\partial_t^{(0)} \nabla p&=&-\nabla \left(\frac{2}{d}a
P_{xy}^{(0)}+p\zeta^{(0)}\right)
\nonumber\\
&=&-\left(\frac{2a}{d}\frac{\partial P_{xy}^{(0)}}{\partial p}+ 2\zeta^{(0)}\right) \nabla
p-\left(\frac{2a}{d}\frac{\partial P_{xy}^{(0)}}{\partial T}-\frac{1}{2} \frac{p\zeta^{(0)}}{T}\right) \nabla T,
\end{eqnarray}
\begin{eqnarray}
\label{b17}
\partial_t^{(0)} \nabla T&=&-\nabla \left(\frac{2T}{d p}a
P_{xy}^{(0)}+\zeta^{(0)}\right)
\nonumber\\
&=&\left[\frac{2aT}{dp^2}\left(1-p\frac{\partial}{\partial p}\right)P_{xy}^{(0)}-\frac{T\zeta^{(0)}}{p}\right]
\nabla
p\nonumber\\
& & -\left[\frac{2a}{dp}\left(1+T\frac{\partial}{\partial T}\right)P_{xy}^{(0)}+\frac{1}{2}\zeta^{(0)}\right]
\nabla T,
\end{eqnarray}
\begin{equation}
\label{b18}
\partial_t^{(0)} \nabla_i \delta u_j=\nabla_i \partial_t^{(0)} \delta u_j=-a_{jk} \nabla_i \delta u_k.
\end{equation}
The corresponding integral equations (\ref{3.19})--(\ref{3.23}) can be obtained when one identifies coefficients
of independent gradients in (\ref{b8}) and takes into account Eqs.\ (\ref{b16})--(\ref{b18}) and the
mathematical property
\begin{eqnarray}
\label{b19}
\partial_t^{(0)} X &=&\frac{\partial X}{\partial p}\partial_t^{(0)}
p+\frac{\partial X}{\partial T}\partial_t^{(0)} T+\frac{\partial X}{\partial \delta u_i}\partial_t^{(0)}
\delta u_i \nonumber\\
&=&-\left(\frac{2}{d p}a P_{xy}^{(0)}+\zeta^{(0)}\right)\left(p\frac{\partial}{\partial
p}+T\frac{\partial}{\partial T}\right) X+a_{ij}\delta u_j \frac{\partial X}{\partial c_i},
\end{eqnarray}
where in the last step we have taken into account that $X$ depends on $\delta {\bf u}$ through ${\bf c}={\bf
V}-\delta {\bf u}$.

\end{document}